%% file: BarfussEtAlInPrep_DetRL_a.tex
\documentclass[%
reprint,
superscriptaddress,
amsmath,amssymb,
aps,
longbibliography,
]{revtex4-1}

\usepackage{dsfont}
\usepackage{mathtools}
\usepackage{caption}
\usepackage[thinlines]{easytable}
\usepackage[overload]{empheq}
\usepackage{graphicx}
\usepackage{dcolumn}
\usepackage{bm}

\usepackage{subcaption}

\begin{document}

\title{Deterministic limit of temporal difference reinforcement learning for stochastic games}

\author{Wolfram Barfuss}
	\email{barfuss@pik-potsdam.de}
	\affiliation{Potsdam Institute for Climate Impact Research, Germany}	\affiliation{Department of Physics, Humboldt University Berlin, Germany}%

\author{Jonathan F. Donges}%
	\affiliation{Potsdam Institute for Climate Impact Research, Germany}%
	\affiliation{Stockholm Resilience Centre, Stockholm University, Sweden}%

\author{J\"urgen Kurths}
	\affiliation{Potsdam Institute for Climate Impact Research, Germany}	\affiliation{Department of Physics, Humboldt University Berlin, Germany}%
	\affiliation{Saratov State University, Russia}%
	
\date{\today}

\begin{abstract}
Reinforcement learning in multiagent systems has been studied in the fields of economic game theory, artificial intelligence and statistical physics by developing an analytical understanding of the learning dynamics (often in relation to the replicator dynamics of evolutionary game theory).
However, the majority of these analytical studies focuses on repeated normal form games, which only have a single environmental state. 
Environmental dynamics, i.e., changes in the state of an environment affecting the agents' payoffs has received less attention, lacking a universal method to obtain deterministic equations from established multistate reinforcement learning algorithms.

In this work we present a novel methodological extension, separating the interaction from the adaptation time scale, to derive the deterministic limit of a general class of reinforcement learning algorithms, called temporal difference learning. This form of learning is equipped to function in more realistic multistate environments by using the estimated value of future environmental states to adapt the agent's behavior.
We demonstrate the potential of our method with the three well established learning algorithms Q learning, SARSA learning and Actor-Critic learning.
Illustrations of their dynamics on two multiagent, multistate environments reveal a wide range of different dynamical regimes, such as convergence to fixed points, limit cycles, and even deterministic chaos.
\end{abstract}

\maketitle

\section{Introduction}
Individual learning through reinforcements is a central approach in the fields of artificial intelligence \cite{SuttonBarto1998,BusoniuEtAl2008,WieringVanOtterlo2012}, neuroscience \cite{Shah2012,HassabisEtAl2017}, learning in games \cite{FudenbergLevine1998} and behavioral game theory \cite{RothErev1995,ErevRoth1998,CamererHo1999,Camerer2003}, thereby offering a general purpose principle to either solve complex problems or explain behavior. 
Also in the fields of complexity economics \cite{Arthur1991,Arthur1999} and social science \cite{MacyFlache2002}, reinforcement learning has been used as a model for human behavior to study social dilemmas.

However, there is a need for improved understanding and better qualitative insight into the characteristic dynamics that different learning algorithms produce. Therefore, reinforcement learning has also been studied from a dynamical systems perspective.
In their seminal work, B\"orgers and Sarin showed that one of the most basic reinforcement learning update schemes, Cross learning \cite{Cross1973}, converges to the replicator dynamics of evolutionary games theory in the continuous time limit \cite{BoergersSarin1997}. 
This has led to at least two, presumably nonoverlapping research communities, one from statistical physics
\cite{MarsiliEtAl2000,SatoEtAl2002,SatoCrutchfield2003,SatoEtAl2005,Galla2009,Galla2011,BladonGalla2011,RealpeGomezEtAl2012,SandersEtAl2012,GallaFarmer2013,AloricEtAl2016}
and one from computer science machine learning
\cite{TuylsEtAl2003,BloembergenEtAl2015,TuylsNowe2005,TuylsEtAl2006,TuylsParsons2007,KaisersTuyls2010,HennesEtAl2009,VrancxEtAl2008,HennesEtAl2010}.
Thus, Sato and Crutchfield \cite{SatoCrutchfield2003} and Tuyls \textit{et al.} \cite{TuylsEtAl2003} independently deduced identical learning equations in 2003.

The statistical physics articles usually consider the deterministic limit of the stochastic learning equations, assuming infinitely many interactions between the agents before an adaptation of behavior occurs. 
This limit can either be performed in continuous time with differential equations \cite{SatoEtAl2002,SatoCrutchfield2003,SatoEtAl2005} or discrete time with difference equations \cite{Galla2009,Galla2011,BladonGalla2011}. The differences between both variants can be significant \cite{Galla2011,RealpeGomezEtAl2012}.
Deterministic chaos was found to emerge when learning simple \cite{SatoEtAl2002} as well as complicated games \cite{GallaFarmer2013}. 
Relaxing the assumption of infinitely many interactions between behavior updates revealed that noise can change the attractor of the learning dynamics significantly, e.g., by noise-induced oscillations \cite{Galla2009,Galla2011}. 

However, these statistical physics studies so far considered only repeated normal form games. These are games where the payoff depends solely on the set of current actions, typically encoded in the entries of a payoff matrix (for the typical case of two players). Receiving payoff and choosing another set of joint actions is performed repeatedly.
This setup lacks the possibility to study dynamically changing environments and their interplay with multiple agents. In those systems, rewards depend not only on the joint action of agents but also on the states of the environment. Environmental state changes may occur probabilistically and depend also on joint actions and the current state. Such a setting is also known as a Markov game or stochastic game \cite{Shapley1953,MertensNeyman1981}. Thus, a repeated normal form game is a special case of a stochastic game with only one environmental state. 
Notably, Akiyama and Kaneko \cite{AkiyamaKaneko2000,AkiyamaKaneko2002} did emphasize the importance of a dynamically changing environment; however did not utilize a reinforcement learning update scheme.

The computer science machine-learning community dealing with reinforcement learning as a dynamical system (see Ref. \cite{BloembergenEtAl2015} for an overview) particularly emphasizes the link between evolutionary game theory and multiagent reinforcement learning as a well grounded theoretical framework for the latter \cite{BloembergenEtAl2015,TuylsNowe2005,TuylsEtAl2006,TuylsParsons2007}.
This dynamical systems perspective is proposed as a way to gain qualitative insights about the variety  of multiagent reinforcement learning algorithms (see Ref. \cite{BusoniuEtAl2008} for a review).
Consequently, this literature developed a focus on the translation of established reinforcement learning algorithms to a dynamical systems description, as well as the development of new algorithms based on insights of a dynamical systems perspective. While there is more work on stateless games (e.g., Q learning \cite{TuylsEtAl2003} and frequency-adjusted multiagent Q learning \cite{KaisersTuyls2010}), multiagent learning dynamics for multistate environments have been developed as well, such as piecewise replicator dynamics \cite{VrancxEtAl2008}, state-coupled replicator dynamics \cite{HennesEtAl2009} or reverse engineering state-coupled replicator dynamics \cite{HennesEtAl2010}.

Both communities, statistical physics and machine learning, share the interest in better qualitative insights into multiagent learning dynamics.
While the statistical physics community focuses more on dynamical properties the same set of learning equations can produce, it leaves a research gap of learning equations capable of handling multiple environmental states.
The machine-learning community, on the other hand, aims more toward algorithm development, but so far has put their focus less on a 
dynamical systems understanding. 
Taken together, there is the challenge of developing a dynamical systems theory of multiagent learning dynamics in varying environmental states. 

With this work, we aim to contribute to such a dynamical systems theory 
of multiagent learning dynamics. We present a novel methodological extension for obtaining the deterministic limit of multistate temporal difference reinforcement learning. In essence, it consists of formulating the temporal difference error for batch learning, and sending the batch size to infinity. 
We showcase our approach with the three prominent learning variants of Q learning, SARSA learning and Actor-Critic learning. 
Illustrations of their learning dynamics reveal multiple different dynamical regimes, such as fixed points, periodic orbits, and deterministic chaos.

In Sec. \ref{sec:Preliminaries} we introduce the necessary background and notation.
Section \ref{sec:DeterministicLimit} presents our method to obtain the deterministic limit of temporal difference reinforcement learning
and demonstrates it for multistate Q learning, SARSA learning and Actor-Critic learning. 
We illustrate their learning dynamics for two previously utilized   two-agents two-actions two-states environments in Sec. \ref{sec:EnvironmentExamples}. In Sec. \ref{sec:Discussion} we conclude with a discussion of our work.

\section{Preliminaries}
\label{sec:Preliminaries}

\begin{figure*}
	\centering
	\includegraphics[width=0.8\linewidth]{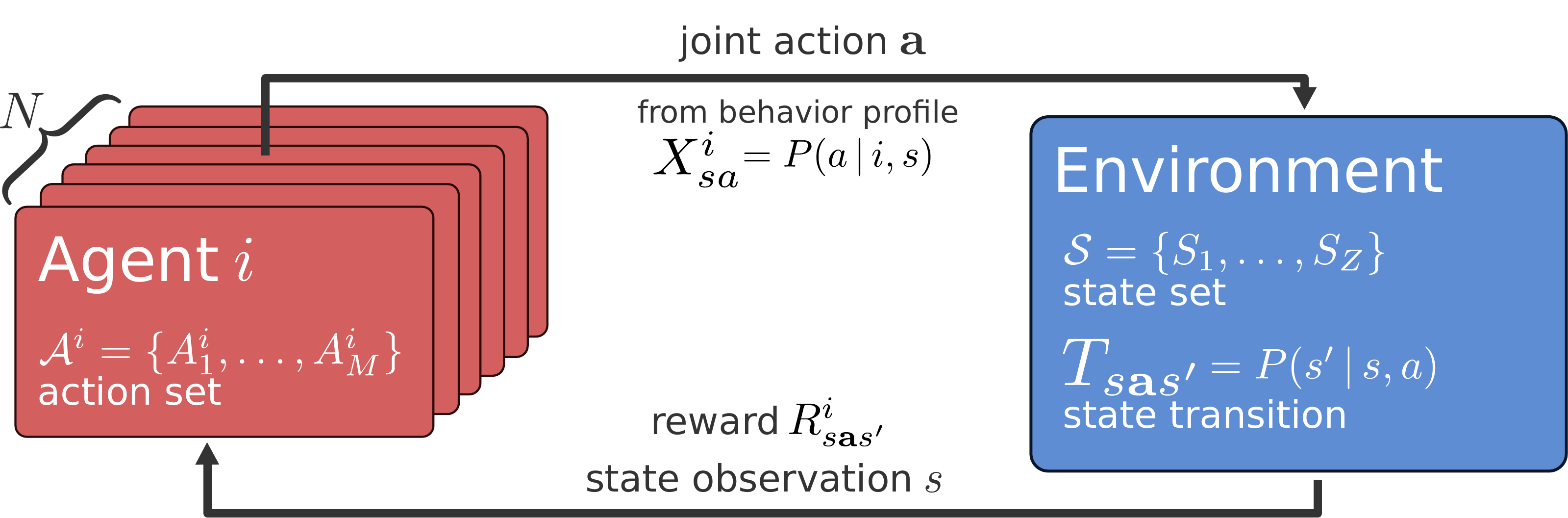}
	\caption{\textbf{Multiagent Markov environment} (also known as stochastic or Markov game). $N$ agents choose a joint action $\mathbf a = (a^1,\dots, a^N)$ from their action sets $\mathcal A^i$, based on the current state of the environment $s$, according to their behavior profile $X^i_{sa} = P(a | i, s)$. This will change the state of the environment from $s$ to $s'$ with probability $T_{s\mathbf a s'}$, and provide each agent with a reward $R^i_{s\mathbf a s'}$.  }
	\label{fig:FrameworkOverview}
\end{figure*}

We introduce the components (including notation) of our multiagent environment systems (see Fig. \ref{fig:FrameworkOverview}), followed by a brief introduction of temporal difference reinforcement learning.

\subsection{Multiagent Markov environments}
\label{sec:Framework-Foundations}

A multiagent Markov environment (also called stochastic game or Markov game) consists of $N \in \mathds{N}$ \textit{agents}.
The environment can exist in $Z \in \mathds{N}$ \textit{states} $\mathcal{S} = \{S_1,\dots, S_Z \}$.
In each state each agent has $M \in \mathds{N}$ available \textit{actions} $\mathcal{A}^i = \{A^i_1,\dots, A^i_M\}, \ i=1,\dots,N$ to choose from.
Having an identical number of actions for all states and all agents is notational convenience, no significant restriction.
A joint action of all agents is referred to by $\mathbf a \in \boldsymbol{\mathcal{A}} = \mathcal{A}^1 \times \cdots \times \mathcal{A}^N$, the joint action of all agents but agent $i$ is denoted by $\mathbf a^{-i} \in \boldsymbol{\mathcal{A}}^{-i} = \mathcal{A}^1 \times \cdots \times \mathcal{A}^{i-1} \times \mathcal{A}^{i+1} \times \cdots \times \mathcal{A}^N$.

Environmental dynamics are given by the probabilities for state changes expressed as a transition tensor 
$
\mathbf{T} \in [0, 1]^{Z \times M \times \dots  \text{($N$ times)} \dots \times M \times Z}.
$
The entry $T_{s\mathbf{a}s'}$ denotes the probability $P(s'|s, \mathbf{a})$ that the environment transitions to state $s'$ given the environment was in state $s$ and the agents have chosen the joint action $\mathbf a$. Hence, for all $s, \mathbf{a}$,  $\sum_{s'} T_{s\mathbf{a}s'} = 1$ must hold. 
The assumption that the next state only depends on the current state and joint action makes our system Markovian. 
We here restrict ourselves to ergodic environments without absorbing states (c.f. Ref. \cite{HennesEtAl2010}). 

The rewards receivable by the agents are given by the reward tensor
$
\mathbf{R} \in \mathds{R}^{N \times Z \times M \times \dots  \text{($N$ times)} \dots \times M \times Z}.
$
The entry $R^i_{s\mathbf{a}s'}$ denotes the reward agent $i$ receives when the environment transits from state $s$ to state $s'$ under the joint action $\mathbf{a}$. Rewards are also called payoffs from a game theoretic perspective.

Agents draw their actions from their behavior profile 
$
\mathbf{X} \in [0, 1]^{N\times Z\times M}.
$
The entry $X^i_{sa} = P(a \,|\, i, s) $ denotes the probability that agent $i$ chooses action $a$ in state $s$.
Thus, for all $i$ and all $s$, $\sum_a X^i_{sa} = 1$ must hold. We here focus on the case of independent agents, able to fully observe the current state of the environment. 
With correlated behavior (see e.g., Ref. \cite{BusoniuEtAl2008}) and partially observable environments \cite{Spaan2012,Oliehoek2012} one could extend the multiagent environment systems to be even more general.
Note that what we call behavior profile is usually termed policy from a machine-learning perspective or behavioral strategy from a game theoretic perspective. We chose to introduce our own term because policies and strategies suggest a deliberate choice which we do not want to impose. 

\subsection{Averaging out behavior and environment}

\newcommand{\AvgA}[1]{\prescript{}{\mathbf X}\!\left\langle {#1} \right\rangle}

\newcommand{\AvgAmI}[1]{\prescript{}{\mathbf X^{-i}}\! \left\langle {#1} \right\rangle}

\newcommand{\AvgSpAmI}[1]{\prescript{}{\mathbf{TX}^{-i}} \!\left\langle {#1} \right\rangle}

\newcommand{\AvgSpA}[1]{\prescript{}{\mathbf{TX}}\!\left\langle {#1} \right\rangle}

\newcommand{\AvgSp}[1]{\prescript{}{\mathbf T}\! \left\langle {#1} \right\rangle}

We define a notational convention, that allows a systematic averaging over the current behavior profile $\mathbf{X}$ and the environmental transitions $\mathbf T$. It will be used throughout the paper.

Averaging over the whole behavioral profile yields
\begin{align}
\AvgA{\circ}
&:= \sum_{\mathbf a} \mathbf X_{s\mathbf a} \cdot \circ \nonumber \\
&:= \sum_{a^1 \in \mathcal A^1} \dots \sum_{a^N \in \mathcal A^N}
X^1_{sa^1} \cdots X^N_{sa^N} \cdot \circ.
\label{eq:AvgA}
\end{align}
Here, $\circ$ serves as a placeholder. If the quantity to be inserted for $\circ$ depends on the summation indices, then those indices will be summed over as well. 
If the quantity, which is averaged out, is used in tensor form, then it is written in bold. If not, then remaining indices are added after the right angle bracket.

Averaging over the behavioral profile of the other agents, keeping the action of agent $i$, yields
\begin{align}
\AvgAmI{\circ}
&:= \sum_{\mathbf a^{-i}} \mathbf X^{-i}_{s\mathbf a^{-i}} \cdot \circ \nonumber \\
& 
:=  \underbrace{\sum_{a^1 \in \mathcal A^1} \dots \sum_{a^N \in \mathcal A^N}}_{\text{excl. } i}
\underbrace{{X^1_{sa^1} \cdots X^N_{sa^N}}}_{\text{excl. } i} 
\cdot \circ.
\label{eq:AvgAmi}
\end{align}

Last, averaging over the subsequent state $s'$ yields
\begin{align}
 \AvgSp{\circ}
& := \sum_{s'} T_{s\mathbf a s'} \cdot \circ := \sum_{s' \in \mathcal S} T_{sa^1\dots a^Ns'} \cdot \circ.
\label{eq:AvgSp}
\end{align}

Of course, these operations may also be combined as $\AvgSpA{\circ}$ and $\AvgSpAmI{\circ}$ by multiplying both summations. 
\newline

For example, given a behavior profile $\mathbf X$, the resulting effective Markov Chain transition matrix reads $\AvgA{T}_{ss'}$, which encodes the transition probabilities from state $s$ to $s'$. From $\AvgA{T}_{ss'}$ the stationary distribution of environmental states $\boldsymbol{\sigma}(\mathbf X)$ can be computed. $\boldsymbol{\sigma}(\mathbf X)$ is the eigenvector corresponding to the eigenvalue 1 of $\AvgA{T}_{ss'}$. Its entries encode the ratios of the average durations the agents find themselves in the respective environmental states.

The average reward agent $i$ receives from state $s$ under action $a$, given all other agents follow the behavior profile $\mathbf X$ reads $\AvgSpAmI{R}^i_{sa}$. 
Including agent $i$'s behavior profile gives the average reward it receives from state $s$: $\AvgSpA{R}^i_{s}$. 
Hence,
$
\AvgSpA{R}^i_{s} = \sum_{a} X^i_{sa} \cdot  \AvgSpAmI{R}^i_{sa}
\label{eq:Rewards-Relation}
$
holds.

\subsection{Agent's preferences and values}
\label{sec:Framework-Preferences}

Typically, agents are assumed to maximize their exponentially discounted sum of future rewards, called return
$
G^i(t) = (1-\gamma^i) \sum_{k=0}^{\infty} (\gamma^i)^k r^i(t+k),
$
where $\gamma^i \in [0, 1)$ is the discount factor of agent $i$ and $r^i(t+k)$ denotes the reward received by agent $i$ at time step $t+k$.
Exponential discounting is most commonly used for its mathematical convenience and because it ensures consistent preferences over time. 
Other formulations of a return use e.g., finite-time horizons, average reward settings, as well as other ways of discounting, such as hyperbolic discounting. Those other forms require their own form of reinforcement learning.

Given a behavior profile $\mathbf X$, the expected return defines the 
\textit{state-value function} $V^{i}_s(\mathbf X) := \prescript{}{\mathbf{TX}}{\left\langle G^i(t) \ | \ s(t) = s \right\rangle^i_s}$
which is independent of time $t$.
The operation $\prescript{}{\mathbf{TX}}{\left\langle \dots | \ s(t) = s  \right\rangle}$ denotes the behavioral and environmental average as defined in Eqs. \ref{eq:AvgA} and \ref{eq:AvgSp} given that in the current time step $t$ the environment is in state $s$.
Inserting the return yields the\textit{ Bellman equation} \cite{Bellman1957}, 
\begin{align}
	V_s^{i} (\textbf{X}) 
	= \prescript{}{\mathbf{TX}}{\left\langle 
		(1-\gamma^i)r^i(t)
	+ \gamma^i V^i_{s(t+1)}(\textbf{X}) \ | \ s(t) = s
	\right\rangle}^i_s.
\label{eq:Bellman}
\end{align}
 
This recursive relationship between state values declares that the value of a state $s$ is the discounted value of the subsequent state $s(t+1)$ plus $(1-\gamma^i)$ times the reward received along the way. Evaluating the behavioral and environmental average $\prescript{}{\mathbf{TX}}{\left\langle \ \right\rangle}$ and writing in matrix form we get:

\begin{equation}
\mathbf V^{i}(\mathbf X) =
	(1-\gamma^i) \cdot \AvgSpA{\mathbf R}^i
	+ \gamma^i \cdot  \AvgA{\mathbf T} \cdot \mathbf V^{i}(\mathbf X).
\label{eq:StateValue-computation-matrixform}
\end{equation}
The reward $r^i(t)$ received at time step $t$ is evaluated to reward $\AvgSpA{ R}^i_s$ for state $s$, since the behavioral and environmental average was conditioned on starting in state $s(t)=s$.
The average subsequent state value $V^i_{s(t+1)}(\textbf{X})$ from the current state $s$ can be expressed as a matrix multiplication of the effective Markov transition matrix and the vector of state values: $\sum_{s'} 
\AvgA{ T}_{ss'} \cdot \mathbf V^{i}_{s'}(\mathbf X)$.

A solution of the state values $\mathbf V^i(\mathbf X)$ can be obtained using matrix inversion
\begin{equation}
\mathbf V^{i}(\mathbf X) 
	= (1-\gamma^i)
	\left(
		\mathds{1}_Z -\gamma^i \AvgA{\mathbf T} 
	\right)^{-1}
	\AvgSpA{\mathbf R}^i.
\label{eq:StateValueComputation}
\end{equation}
The computational complexity of matrix inversion makes this solution strategy infeasible for large systems. Therefore many iterative solution methods exist \cite{WieringVanOtterlo2012}.

Equivalently, \textit{state-action-value functions} $Q^i_{sa}$ are defined as the expected return, given agent $i$ applied action $a$ in state $s$ and then followed $\mathbf X$ accordingly:
$
Q^{i}_{sa}(\textbf{X}) :=
\prescript{}{\mathbf{TX}}{\left\langle 
G^i(t) \ | \ s(t) = s, a(t) = a
\right\rangle}^i_{sa}.
$ Even though this is the behavioral average over the whole behavioral profile, the resulting object carries an action index because the operation is conditioned on the current action to be $a(t) =a$.
They can be computed via
\begin{align}
Q^i_{sa}(\mathbf X) = 
(1-\gamma^i) \AvgSpAmI{R}^i_{sa} 
+ \gamma^i\sum_{s'}  \AvgA{T}_{ss'} \cdot V^i_{s'}(\mathbf X).
\label{eq:StateActionValueComputation}
\end{align}

One can show that
$
V^i_s(\mathbf X) = \sum_a X^i_{sa} Q^i_{sa}(\mathbf X)
$
holds for the inverse relation of state-action and state values.

\subsection{Learning through reinforcement}
\label{sec:Framework-ReinforcementLearning}

In contrast to the typical game-theoretic assumption of perfect information, we assume that agents know nothing about the game in advance. They can only gain information about the environment and other agents through interactions. 
They do not know the true reward tensor $\mathbf R$ or the true transition probabilities $T_{s\mathbf a s'}$.
They experience only reinforcements (i.e., particular rewards $R^i_{s\mathbf a s'}$), while observing the current true Markov state of the environment.  

In essence, reinforcement learning consists of iterative behavior changes toward a behavior profile with maximum state values.
However, due to the agents' limited information about the environment, they generally cannot compute a behavior profile's true state- and state-action values, $V^i_{s}(\mathbf X)$ and $Q^i_{sa}(\mathbf X)$, as defined in the previous section.
Therefore, agents use time-dependent \textit{state-value} and \textit{state-action-value approximations}, $\tilde V^i_s(t)$ and $\tilde Q^i_{sa}(t)$, during the reinforcement learning process.

\subsubsection{Temporal difference learning}

Basically, state-action-value approximations $\tilde Q^i_{sa}$ get iteratively updated by a temporal difference error $T\!D^i_{sa}(t)$:
\begin{equation}
\tilde Q^i_{sa}(t+1) = \tilde Q^i_{sa}(t) + \alpha^i T\!D^i_{sa}(t),
\label{eq:QUpdate}
\end{equation}
with $\alpha^i \in (0, 1)$ being the \textit{learning rate} of agent $i$.
These state-action propensities $\tilde Q^i_{sa}$ can be interpreted as estimates of the state-action values $Q^i_{sa}$.

The temporal difference error expresses a difference in the estimation of state-action values. 
New experience is used to compute a new estimate of the current state-action value and corrected by the old estimate. 
The estimate from the new experience uses exactly the recursive relation of value functions from the Bellmann equation (Eq. \ref{eq:Bellman}),

\begin{align}
& T\!D^{i}_{sa}(t) = \delta_{ss(t)}\delta_{aa(t)} \cdot \nonumber \\
& \quad \Bigg[ 
\underbrace{(1-\gamma^i)  R^i_{s(t)a(t)\mathbf{a}^{-i}(t)s(t+1)}
	+ \gamma^i \curlyvee^i_{s(t+1)}(t)
}_\text{estimate from new experience} \nonumber \\
& \qquad - 
\underbrace{\curlyvee^i_{s(t)}(t)
}_\text{old estimate}
\Bigg].
\label{eq:TemporalDifferenceError}
\end{align}
Here $s$ and $a$ denote the state-action pair whose temporal difference error is calculated. With $s(t)$, $a(t)$, etc. we refer to the state, action, etc. that occurred 
at time step $t$. Thus, the notation $R^i_{s(t)a(t)\mathbf{a}^{-i}(t)s(t+1)}$ refers to the entry of the reward tensor $R^i_{sa\mathbf a^{-i}s'}$ when at time step $t$ the environmental state was $s$ [$s(t)=s$], agent $i$ chose action $a$ [$a(t)=a$], the other agents chose the joint action $\mathbf a^{-i}$ [$\mathbf a^{-i}(t) = \mathbf a^{-i}$] and the next environmental state was $s'$ [$s(t+1) = s'$].
The $\curlyvee^i_{s(t+1)}(t)$ indicates the state-value estimate at time step $t$  of the state visited at the next time step $s(t+1)$. $\curlyvee^i_{s(t)}(t)$ denotes the state-value estimate at time step $t$ of the current state $s(t)$. Different choices for these estimations are possible, leading to different learning variants (see below). 

The Kronecker deltas $\delta_{ss(t)},\delta_{aa(t)}$ indicate that the temporal difference error for state-action pair $(s, a)$ is only nonzero when $(s, a)$ was actually visited in time step $t$. This denotes and emphasizes that agents can only learn from experience. In contrast, e.g., experience-weighted-attraction learning \cite{CamererHo1999} assumes that action propensities can be updated with hypothetical rewards an agent would have received if they had played a different action than the current action.
These two cases have been referred to as \textit{full vs. partial information} \cite{MarsiliEtAl2000}. Thus, the Kronecker deltas in Eq. \ref{eq:TemporalDifferenceError} indicate a partial information update. The agents use only information experienced through interaction.

The state-action-value approximations $\tilde Q^i_{sa}$ are translated to a behavior profile according to the
Gibbs-Boltzmann distribution \cite{SuttonBarto1998} (also called softmax)
\begin{equation}
X^i_{sa}(t) = \frac{ \exp(\beta^i \tilde {Q}^i_{sa}(t) )}{\sum_b \exp(\beta^i \tilde Q^i_{sb}(t))}.
\label{eq:BoltzmannBehavior}
\end{equation}
The behavior profile $\mathbf X$ becomes a dynamic variable as well.
The parameter $\beta^i$ controls the \textit{intensity of choice} or the \textit{exploitation level} of agent $i$ controlling the \textit{exploration-exploitation trade-off}.
In analogy to statistical physics, $\beta^i$ is the inverse temperature.
For high $\beta^i$ agents tend to exploit their learned knowledge about the environment, leaning toward actions with high estimated state-action value.
For low $\beta^i$, agents are more likely to deviate from these high value actions in order to explore the environment further with the chance of finding actions, which eventually lead to even higher values.
Other behavior profile translations exist as well (e.g., $\epsilon$-greedy \cite{SuttonBarto1998}).

\subsubsection{Three learning variants}
The specific choices of the value estimates $\curlyvee$ in the temporal difference error result in different reinforcement learning variants. \newline

\paragraph{Q learning.}
For the Q learning algorithm \cite{SuttonBarto1998,WieringVanOtterlo2012}, $\curlyvee^i_{s(t+1)}(t) = \max_b \tilde Q^i_{s(t+1)b}(t)$ and $\curlyvee^i_{s(t)}(t) = \tilde Q^i_{s(t)a(t)}(t)$.
Thus, the Q learning update takes the maximum of the next state-action-value approximations as an estimate for the next state value, regardless of the actual next action the agent plays. This is reasonable because the maximum is the highest value achievable given the current knowledge. 
For the state-value estimate of the current state, the Q learner takes the current state-action-value approximation $Q^i_{s(t)a(t)}(t)$. This is reasonable because it is exactly the quantity that gets updated by Eq. \ref{eq:QUpdate}.

\paragraph{SARSA learning.}
For SARSA learning \cite{SuttonBarto1998,WieringVanOtterlo2012}, $\curlyvee^i_{s(t+1)}(t) = \tilde Q^i_{s(t+1)a(t+1)}(t)$ and $\curlyvee^i_{s(t)}(t) = \tilde Q^i_{s(t)a(t)}(t)$, where $a(t+1)$ denotes the action taken by agent $i$ at the next time step $a(t+1) = A^i(t+1)$.
Thus, the SARSA algorithm uses the five ingredients of an update sequence of State, Action, Reward, next State, next Action to perform one update. 
In practice, the SARSA sequence has to be shifted one time step backward to know what the actual "next" action of the agent was.

\paragraph{Actor-Critic (AC) learning.}
For AC learning \cite{SuttonBarto1998,WieringVanOtterlo2012}, $\curlyvee^i_{s(t+1)}(t) = \tilde V^i_{s(t+1)}(t)$ and $\curlyvee^i_{s(t)}(t) = \tilde V^i_{s(t)}(t)$. Compared to Q and SARSA learners, it has an additional data structure of state-value approximations which get separately updated according to
$
\tilde V^i_s(t+1) = \tilde V^i_s(t) + \alpha^i \cdot D^{i}_{sa}(t).
$
The state-action-value approximations $\tilde Q^i_{sa}$ serve as the actor which gets criticized by the state-value approximations $\tilde V^i_s$.

Tab. \ref{tab:LEO_K1} summarizes the values estimates $\curlyvee$ for these three learning variants. Q and SARSA learning are structurally more similar compared to the Actor-Critic learner, which uses an additional data structure of state-value approximations $\tilde V^i_s$.

\renewcommand{\arraystretch}{2}
\begin{table*}[]
	\begin{subtable}[h]{0.3\textwidth}
		\centering
		\begin{tabular}{l}
			 \\ \hline\hline
			Q learning \\ \hline
			SARSA learning \\ \hline
			Actor Critic (AC) learning 	
		\end{tabular}	
		\vspace{0.6cm}
	\end{subtable}
	\begin{subtable}[h]{0.27\textwidth}
		\centering
		\begin{tabular}{cccc}
			$\curlyvee^i_{s(t+1)}(t)$ & $\curlyvee^i_{s(t)}(t)$ \\
			\hline\hline
			\ $\max_b \tilde Q^i_{s(t+1)b}(t)$   \
			& \ $\tilde Q^i_{s(t)a(t)}(t)$ \ \\ \hline
			$\tilde Q^i_{s(t+1)a(t+1)}(t)$  
			& $\tilde Q^i_{s(t)a(t)}(t)$ \\ \hline
			$\tilde V^i_{s(t+1)}(t)$ 
			& $\tilde V^i_{s(t)}(t)$ 
		\end{tabular}	
		\caption{$K=1$}
		\label{tab:LEO_K1}
	\end{subtable}
	\begin{subtable}[h]{0.27\textwidth}
		\centering
		\begin{tabular}{cccc}
			$\curlyvee^i_{s(t+1)}(t)$ & $\curlyvee^i_{s(t)}(t)$ \\
			\hline\hline
			\ $\prescript{\text{max}}{}{\mathcal Q}^i_{sa}(\mathbf X)$   \
			& \ $\frac{1}{\beta^i}\log X^i_{sa} (t)$ \ \\ \hline
			$\prescript{\text{next}}{}{\mathcal V}^{i}_{sa}(\mathbf X)$  
			& $\frac{1}{\beta^i}\log X^i_{sa} (t)$ \\ \hline
			$\prescript{\text{next}}{}{\mathcal V}^{i}_{sa}(\mathbf X)$ 
			& / 
		\end{tabular}	
		\caption{$K=\infty$}
		\label{tab:LEO_Kinfty}
	\end{subtable}
	\caption{Overview of the three reinforcement learning variants. Shown in the columns are the value estimates for the next state $\curlyvee^i_{s(t+1)}(t)$ and the current state $\curlyvee^i_{s(t)}(t)$ for both ends of the batch size spectrum: $K=1$ and $K=\infty$.}
	\label{tab:LearingEquationsOverview}
\end{table*}

\section{Deterministic limit}
\label{sec:DeterministicLimit}

So far we gave a brief introduction to temporal difference reinforcement learning. A more comprehensive presentation can be found in Ref. \cite{SuttonBarto1998}. 
In this section we will present a novel extension to the methodology of interaction-adaptation timescales separation to the general class of temporal difference reinforcement learning.
In summary, we (i) give a batch formulation of the temporal difference error, (ii) separate the timescales of interaction and adaptation by sending the batch size to infinity and (iii) present a resulting deterministic limit conversion rule for discrete time updates.
We showcase our method in the three learning variants of Q, SARSA and Actor-Critic learning. 
For the statistical physics community, the novelty consists of learning equations, capable of handling environmental state transitions. For the machine learning community the novelty lies in the systematic methodology we use to obtain the deterministic learning equations.  
Note that these deterministic learning equations will not depend on the state-value or state-action-value approximations anymore, being iterated maps of the behavior profile alone. 

Following e.g., Refs. \cite{SatoCrutchfield2003,SatoEtAl2005,BladonGalla2011}, we first combine Eqs.  \ref{eq:QUpdate} and \ref{eq:BoltzmannBehavior} and obtain
\begin{align}
X^i_{sa}(t+1)
&= \frac{X^i_{sa}(t)  \exp\Big(\alpha^i\beta^i T\!D^i_{sa}(t)\Big)}
{\sum_b X^i_{sb}(t)  \exp\Big(\alpha^i\beta^i T\!D^i_{sb}(t)\Big)}.
\label{eq:JointPolicyIteration} 
\end{align}
Although it appears that only the product $\alpha^i\beta^i$ matters for a behavior profile update, the temporal difference error $T\!D^i_{sa}$ may depend only on the exploitation level $\beta^i$, as we will show below.

Next, we formulate the temporal difference error for batch learning.

\subsection{Batch learning}
With batch learning we mean that several time steps of interaction with the environment and the other agents take place before an update of the state-action-value approximations and the behavior profile occurs. 
It has also been interpreted as a form of history replay \cite{LangeEtAl2012} which is essential to stabilize the learning process when function approximation (e.g., by deep neural networks) is used \cite{MnihEtAl2015}. History (i.e., already experienced state, action, next state triples) is used again for an update of the state-action-value approximations.

Imagine that the information from these interactions are stored inside a batch of size  $K \in \mathds{N}$. We introduce the corresponding temporal difference error of batch size $K$:
\begin{align}
&  T\!D_{sa}^{i}(t; K) := \frac{1}{K(s,a)} \sum_{k=0}^{K-1}
  \Big[
  \delta_{ss(t+k)}\delta_{aa(t+k)} \nonumber \\
  & \quad\qquad 
  \cdot 
  \Big(
  (1-\gamma^i)  R^i_{s(t+k)a(t+k)\mathbf{a}^{-i}(t+k)s(t+k+1)} \nonumber \\
  & \quad\qquad \quad + 
  \gamma^i \curlyvee^i_{s(t+k+1)}(t) - \curlyvee^i_{s(t)}(t)
  \Big)
  \Big]
\end{align}
where $K(s,a) = \max(1, \sum_{k=0}^{K-1} \delta_{ss(t+k)}\delta_{aa(t+k)})$ denotes the number of times the state-action pair $(s,a)$ was visited. If the state-action pair $(s,a)$ was never visited, then $K(s,a)=1$.
The agents interact $K$ times under the same behavior profile and use 
the sample average 
to summarize the new experience in order to update the state-action-value approximations:
\begin{equation}
\tilde Q^i_{sa}(t+K) = \tilde Q^i_{sa}(t) + \alpha^i T\!D^{i}_{sa}(t; K). 
\end{equation}
The notation $T\!D^i_{sa}(t)$ denotes a batch update of batch size 1: $T\!D^i_{sa}(t) = T\!D^i_{sa}(t; 1)$.

\subsection{Separation of timescales}

We obtain the deterministic limit of the temporal difference learning dynamics by sending the batch size to infinity, $K \rightarrow \infty$.
Equivalently, this can be regarded as a separation of timescales. 
Two processes can be distinguished during an update of the state-action-value approximations $\Delta \tilde Q^i_{sa}(t) :=
\tilde Q^i_{sa}(t+1) - \tilde Q^i_{sa}(t)$: adaptation and interaction, 
\begin{align}
& \Delta \tilde Q^i_{sa}(t) =
\alpha^i \delta_{ss(t)}\delta_{aa(t)}  \cdot \nonumber \\
& \ \overbrace{
	 \Big(
	\underbrace{(1-\gamma^i) R^i_{s(t)a(t)\mathbf{a}^{-i}(t)s(t+1)} + \gamma^i \curlyvee^i_{s(t+1)}(t)}_{\text{interaction}} - \curlyvee^i_{s(t)}(t)
	\Big)
}^{\text{adaptation}}
\label{eq:Qupdate-timescales}
\end{align}
By separating the timescales of both processes, we assume that (infinitely) many interactions happen before one step of behavior profile adaptation occurs. 

Under this assumption and because of the assumed ergodicity one can replace the sample average, i.e., the sum over sequences of states and actions with the behavior profile average, i.e., the sum over state-action behavior and transition probabilities according to
\begin{equation}
\boxed{
\frac{1}{K(s,a)} \sum_{k=0}^{K-1}
\delta_{ss(t+k)}\delta_{aa(t+k)} \rightarrow
\sum_{s'} \sum_{\mathbf a^{-i}}  \mathbf X^{-i}_{s\mathbf a^{-i}} T_{sa \mathbf a^{-i} s'}.
\label{eq:BatchUpdate}
}
\end{equation}

For example, the immediate reward $R^i_{s(t)a(t)\mathbf{a^{-i}}(t)s(t+1)}$ in the temporal difference error becomes $\AvgSpAmI{R}^i_{sa}$. The time $t$ gets resealed accordingly, as well. 

Taking the limit $K \rightarrow \infty$ in this way, we choose to stay in discrete time, leaving the continuous time limit following Refs.
\cite{SatoCrutchfield2003,SatoEtAl2005,GallaFarmer2013} for future work.

\subsection{Three learning variants}
Next, we present the deterministic limit of the temporal difference error of the three learning variants of Q, SARSA and Actor-Critic learning. Inserting them into Eq. \ref{eq:JointPolicyIteration} yields the complete description of the behavior profile update in the deterministic limit. Table \ref{tab:LearingEquationsOverview} presents an overview of the resulting equations and a comparison to their batch size $K=1$ versions.

\subsubsection{Q learning}
The temporal difference error of Q learning consists of three terms: (i) $R^i_{s(t)a(t)\mathbf{a^{-i}}(t)s(t+1)}$, (ii) $\max_b \tilde Q^i_{s(t+1)b}(t)$, and (iii) $\tilde Q^i_{s(t)a(t)}(t)$. 
As already stated, $R^i_{s(t)a(t)\mathbf{a^{-i}}(t)s(t+1)} \rightarrow \AvgSpAmI{R}^i_{sa}$ under $K\rightarrow \infty$.  $\max_b \tilde Q^i_{s(t+1)b}(t) \rightarrow \prescript{\text{max}}{}{\mathcal Q}^i_{sa}(\mathbf X)$, which is defined as
\begin{equation}
\prescript{\text{max}}{}{\mathcal Q}^i_{sa}(\mathbf X) := \sum_{s'} \sum_{\mathbf a^{-i}}  \mathbf X^{-i}_{s\mathbf a^{-i}} T_{sa \mathbf a^{-i} s'} \max_b Q^i_{s'b}(\mathbf X)
\label{eq:AnaMaxQ}
\end{equation}
using the deterministic limit conversion rule (Eq. \ref{eq:BatchUpdate}).
Because of the assumption of infinite interactions, we can here replace the state-action-value approximations $\tilde Q^i_{s(t+1)b}$ with the true state-action values $Q^i_{s'b}$ as defined by Eq. \ref{eq:StateActionValueComputation}. 

For the third term, we invert Eq. \ref{eq:BoltzmannBehavior}, yielding $\tilde Q^i_{sa}(t) = (\beta^i)^{-1} \log X^i_{sa} (t) + \text{const}^i_s$, where $\text{const}^i_s$ is constant in actions but may vary for each agent and state.
Now, one can show that the dynamics induced by Eq. \ref{eq:JointPolicyIteration} are invariant against additive transformations in the temporal difference error $T\!D^i_{sa}(t, \infty) \rightarrow T\!D^i_{sa}(t, \infty) + \text{const}^i_s$.
Thus, the third term can be converted according to $\tilde Q^i_{s(t)a(t)}(t) \rightarrow (\beta^i)^{-1} \log X^i_{sa} (t)$.

All together, the temporal difference error for Q learning in the deterministic limit reads
\begin{align}
 \prescript{\text{q}}{}T\!D^i_{sa}(t, \infty) = & 
(1-\gamma^i) \AvgSpAmI{R}^i_{sa}
 \nonumber \\
& + \gamma^i \prescript{\text{max}}{}{\mathcal Q}^i_{sa}(\mathbf X)
  - \frac{1}{\beta^i}\log X^i_{sa} (t)
\label{eq:TDe_Q}.
\end{align}

\subsubsection{SARSA learning}
Two of the three terms of the SARSA temporal difference error are identical to the one of Q learning, leaving $ \tilde Q^i_{s(t+1)a(t+1)}(t)$, which we replace by
\begin{equation}
\prescript{\text{next}}{}{\mathcal Q}^i_{sa}(\mathbf X) :=
\sum_{s'} \sum_{\mathbf a^{-i}}  \mathbf X^{-i}_{s\mathbf a^{-i}} T_{sa \mathbf a^{-i} s'} \sum_b X^i_{s'b} Q^i_{s'b} (\mathbf X)
\end{equation}
using again the deterministic limit conversion rule (Eq. \ref{eq:BatchUpdate}) and  the state-action value $Q^i_{s'b}(\mathbf X)$ of the behavior profile $\mathbf X$ according to Eq. \ref{eq:StateActionValueComputation}. 

Thus, the temporal difference error for the SARSA learning update in the deterministic limit reads
\begin{align}
\prescript{\text{sarsa}}{}T\!D^i_{sa}(t; \infty) & = 
(1-\gamma^i) \AvgSpAmI{R}^i_{sa} \nonumber \\
& + \gamma^i \prescript{\text{next}}{}{\mathcal Q}^i_{sa}(\mathbf X)
- \frac{1}{\beta^i}\log X^i_{sa} (t).
\label{eq:TDe_S}
\end{align}

\subsubsection{Actor-Critic (AC) learning}
For the temporal difference error for AC learning we have to find replacements for (i) $ \tilde V^i_{s(t+1)}(t)$ and (ii) $\tilde V^i_{s(t)}(t)$. Applying again Eq. \ref{eq:BatchUpdate} yields $\tilde V^i_{s(t+1)}(t) \rightarrow \prescript{\text{next}}{}{\mathcal V}^{i}_{sa} $ defined as
\begin{equation}
\prescript{\text{next}}{}{\mathcal V}^{i}_{sa} := \sum_{s'} \sum_{\mathbf a^{-i}}  \mathbf X^{-i}_{s\mathbf a^{-i}} T_{sa \mathbf a^{-i} s'} V^{i}_{s'} (\mathbf X),
\end{equation}
using Eq. \ref{eq:StateValueComputation} for the state value $V^{i}_{s'} (\mathbf X)$. This is the average value of the next state given that in the current state the agent took action $a$.
One can show that $\prescript{\text{next}}{}{\mathcal V}^{i}_{sa}(\mathbf X) = \prescript{\text{next}}{}{\mathcal Q}^i_{sa}(\mathbf X)$ from the SARSA update.

The second remaining term belongs to the slower adaptation timescale or, in other words, occurs outside the batch. Thus, our deterministic limit conversion rule (Eq. \ref{eq:BatchUpdate}) does not apply. We could think of a conversion $\tilde V^i_{s(t)}(t) := \sum_a X^i_{sa} \tilde Q^i_{s(t)a(t)}(t) \rightarrow (\beta^i)^{-1} \sum_a X^i_{sa}(t) \log X^i_{sa} (t)$. However, the remaining term is constant in action, and therefore irrelevant for the dynamics, as we have argued above. Thus, we can simply put $\tilde V^i_{s(t)}(t) \rightarrow 0$.

All together, the temporal difference error of the Actor-Critic learner in the deterministic limit reads
\begin{align}
\prescript{\text{ac}}{}T\!D^i_{sa}(t, \infty) = &
(1-\gamma^i) \AvgSpAmI{R}^i_{sa} \nonumber \\
& + \gamma^i \prescript{\text{next}}{}{\mathcal V}^{i}_{sa}(\mathbf X)
\label{eq:TDe_A}
\end{align}

\section{Application to example environments}
\label{sec:EnvironmentExamples}

In the following we apply the derived deterministic learning equations in two different environments. Specifically, we compare the three well established temporal difference learning variants (Q learning, SARSA learning and Actor-Critic (AC) learning) in two different two-agent ($N=2$), two-action ($M=2$) and two-state ($Z=2$) environments: a two-state Matching Pennies game and a two-state Prisoner's Dilemma.
Since the main contribution of this paper is the derivation of the deterministic temporal difference learning equations, we are not trying to make a case with our example environments beyond a systematic comparison of our learners. Therefore, we chose environments that have been used previously in related literature \cite{VrancxEtAl2008,HennesEtAl2009,HennesEtAl2010,HilbeEtAl2018}. 
Note also that we leave a comparison between the deterministic limit and the stochastic equations to future work, which would add a noise term to our equations following the example of Ref. \cite{Galla2009}.

To measure the performance 
of an agent's behavior profile in a single scalar, we use the dot product between the stationary state distribution $\boldsymbol{\sigma}(\mathbf X)$ of the effective Markov Chain with the transition matrix $\AvgA{\mathbf T}$ and the behavior average reward $\AvgSpA{\mathbf R}^i$. Interestingly, we find this relation to be identical to the dot product of the stationary distribution and the state value $\mathbf V^{i}(\mathbf X)$:
\begin{equation}
\boldsymbol{\sigma}(\mathbf X) \cdot \AvgSpA{\mathbf R}^i = \boldsymbol{\sigma}(\mathbf X) \cdot \mathbf V^{i}(\mathbf X). 
\label{eq:Performance}
\end{equation}
This relation can be shown by using Eq. \ref{eq:StateValueComputation} and the fact that $\boldsymbol{\sigma}(\mathbf X)$ is an eigenvector of $\AvgA{\mathbf T}$.

In the following examples we will only investigate homogeneous agents, i.e., agents whose parameters will not differ from each other. We will therefore drop the agent indices from $\alpha^i, \beta^i$, and $\gamma^i$. The heterogeneous agent case is to be explored in future work.

\begin{figure}[ht]
	\centering
	\includegraphics[width=0.98\linewidth]{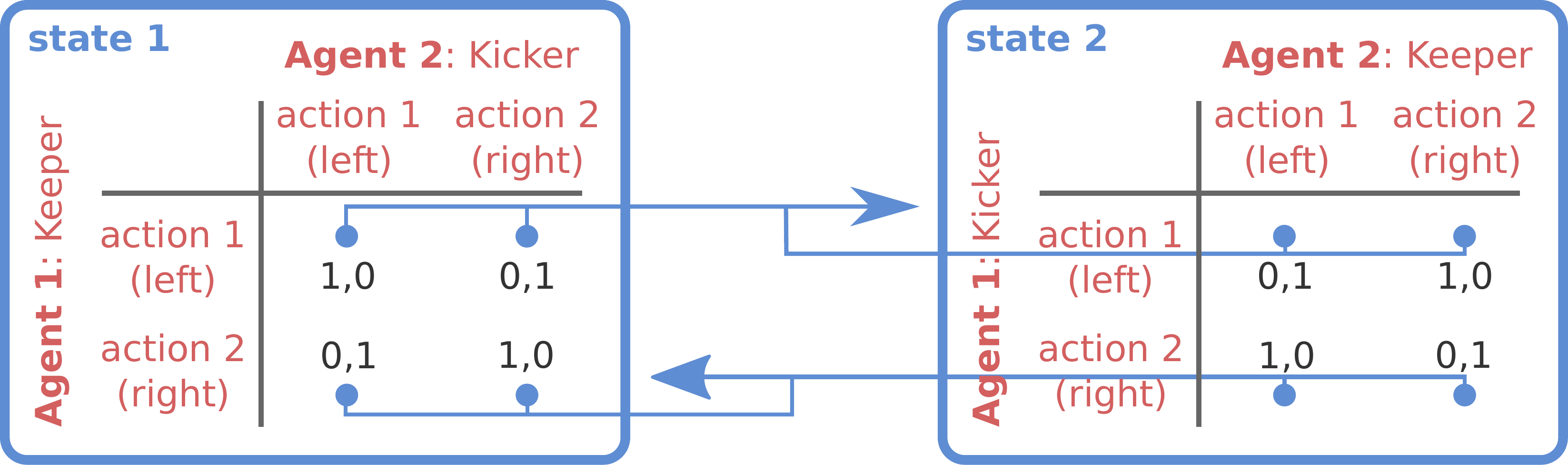}
	\caption{\textbf{Two-state Matching Pennies}. Rewards are given in black type in the payoff tables for each state. State-transition probabilities are indicated by (blue) arrows.}
	\label{fig:MPgame}
\end{figure}

\begin{figure*}
	\centering
	\includegraphics[width=1.0\linewidth]{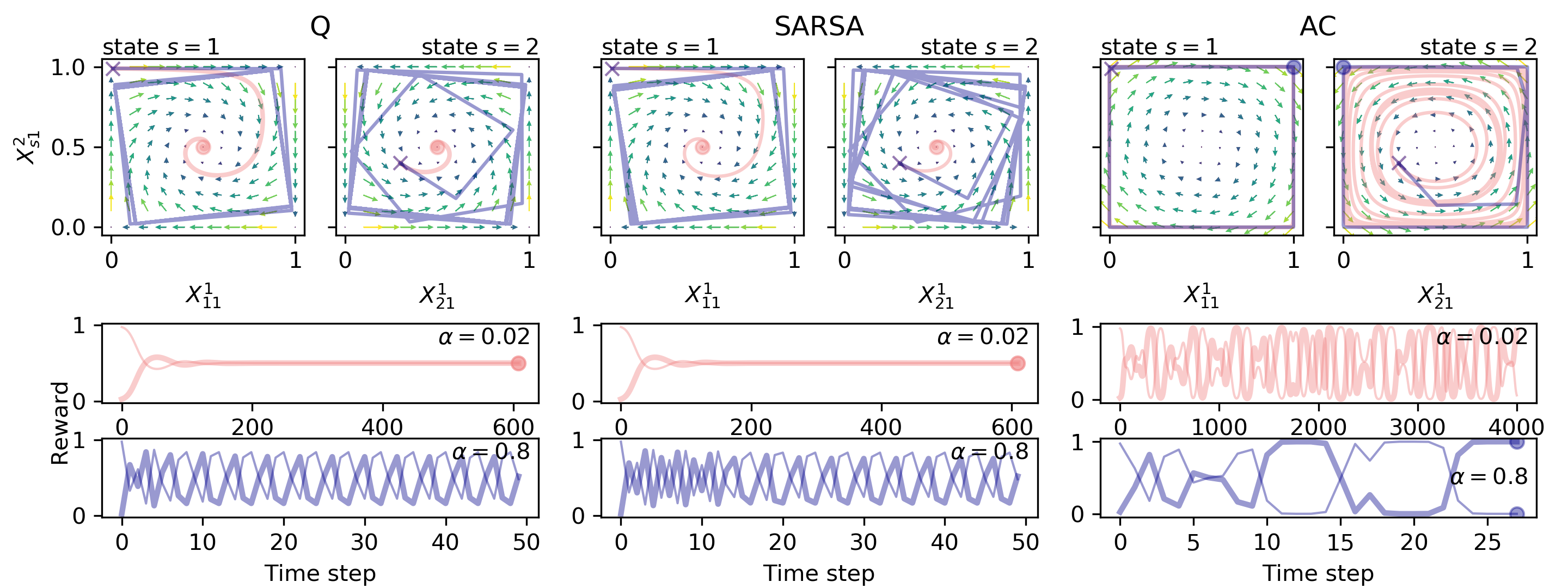}
	\caption{\textbf{Three learners in two-state Matching Pennies environment} for low discount factor $\gamma=0.1$; intensity of choice $\beta=5.0$.
		At the top, the temporal difference errors for the Q learner (Eq. \ref{eq:TDe_Q}), SARSA learner (Eq. \ref{eq:TDe_S} ) and Actor-Critic (AC) learner (Eq. \ref{eq:TDe_A})
		are shown in two behavior phase-space sections, one for each state. The arrows indicate the average direction the temporal difference errors drive the learner toward, averaged over all phase-space points of the other state. Arrow colors (and shadings) additionally encode their lengths. Selected trajectories are shown in the phase-space sections, as well as by reward trajectories, plotting the average reward value (Eq. \ref{eq:Performance}) over time steps. Crosses in the phase-space subsections indicate the initial behavior $(X^1_{11}, X^2_{11}, X^1_{21}, X^2_{21}) = (0.01, 0.99, 0.3, 0.4)$. Circles signal the arrival at a fixed point, determined by the absolute difference of behavior profiles between two subsequent time steps being below $\epsilon=10^{-6}$. Trajectories are shown for two different learning rates $\alpha=0.02$ (light red) and $\alpha=0.8$ (dark blue). The bold reward trajectory belongs to agent 1 and the thin one to agent 2. Note that the temporal difference error is independent from the learning rate $\alpha$. A variety of qualitatively different dynamical regimes can be observed.}
	\label{fig:MP_lowgamma}
\end{figure*}

\subsection{Two-state Matching Pennies}

The single-state matching pennies game is a paradigmatic two-agents, two-actions game. 
Imagine the situation of soccer penalty kicks. The keeper (agent 1) can choose to jump either to the left or right side of the goal, and the kicker (agent 2) can choose to kick the ball also either to the left or the right. If both agents choose the identical side, then the keeper agent wins, otherwise the kicker wins. 

In the two-state version of the game, according to Ref. \cite{HennesEtAl2010}, the rules are extend as follows: In state 1 the situation is as described in the single-state version. Whenever agent 1 (the keeper) decides to jump to the left, the environment transitions to state 2, in which the agents switch roles: Agent 1 now plays the kicker and agent 2 the keeper. From here, whenever agent 1 (now the kicker) decides to kick to the right side the environment transition again to state 1 and both agents switch their roles again.

Figure \ref{fig:MPgame} illustrates this two-state Matching Pennies games. Formally, the payoff matrices are given by
\begin{equation*}
\left(\begin{smallmatrix}
R^1_{111s'}, R^2_{111s'} & R^1_{112s'}, R^2_{112s'} \\
R^1_{121s'}, R^2_{121s'} & R^1_{122s'}, R^2_{122s'}
\end{smallmatrix}\right) 
=
	\left(\begin{smallmatrix}
1,0 & 0,1 \\
0,1 & 1,0
\end{smallmatrix}\right) 
\end{equation*}
in state $1$ and
	\begin{equation*}
	\left(\begin{smallmatrix}
	R^1_{211s'}, R^2_{211s'} & R^1_{212s'}, R^2_{212s'} \\
	R^1_{221s'}, R^2_{221s'} & R^1_{222s'}, R^2_{222s'}
	\end{smallmatrix}\right) 
	=
	\left(\begin{smallmatrix}
	0,1 & 1,0 \\
	1,0 & 0,1
	\end{smallmatrix}\right) 
	\end{equation*}
in state $2$ for $s' \in \{1, 2\}$.
State transitions are governed by
	\begin{equation*}
	\left(\begin{smallmatrix}
	T_{1112} & T_{1122}\\
	T_{1212} & T_{1222}
	\end{smallmatrix}\right) 
	=
	\left(\begin{smallmatrix}
	1 & 1 \\
	0 & 0
	\end{smallmatrix}\right) 
	\ \text{and} \
	\left(\begin{smallmatrix}
	T_{2111} & T_{2121}\\
	T_{2211} & T_{2221}
	\end{smallmatrix}\right) 
	=
	\left(\begin{smallmatrix}
	0 & 0 \\
	1 & 1
\end{smallmatrix}\right). 
	\end{equation*}

Thus, by construction, the probability of transitioning to the other state is independent of agent 2's action. Only agent 1 has agency over the state transitions. 
By playing a uniformly random behavior profile 
$(X^1_{11}, X^2_{11}, X^1_{21}, X^2_{21}) = (0.5, 0.5, 0.5, 0.5)$,
both agents would obtain an average reward of $0.5$ per time step.
 
With Fig. \ref{fig:MP_lowgamma} we compare the temporal difference error in the behavior space sections for each environmental state at a comparable low discount factor $\gamma \in (0, 1)$ of $\gamma=0.1$, as well as learning trajectories for an exemplary initial condition for two learning rates $\alpha \in (0, 1)$, a low one ($\alpha = 0.02$) and a high one ($\alpha = 0.8$).
Overall, we observe a variety of qualitatively different dynamical regimes, such as fixed points, periodic orbits and chaotic motion.

\begin{figure}
	\centering
	\includegraphics[width=.9\linewidth]{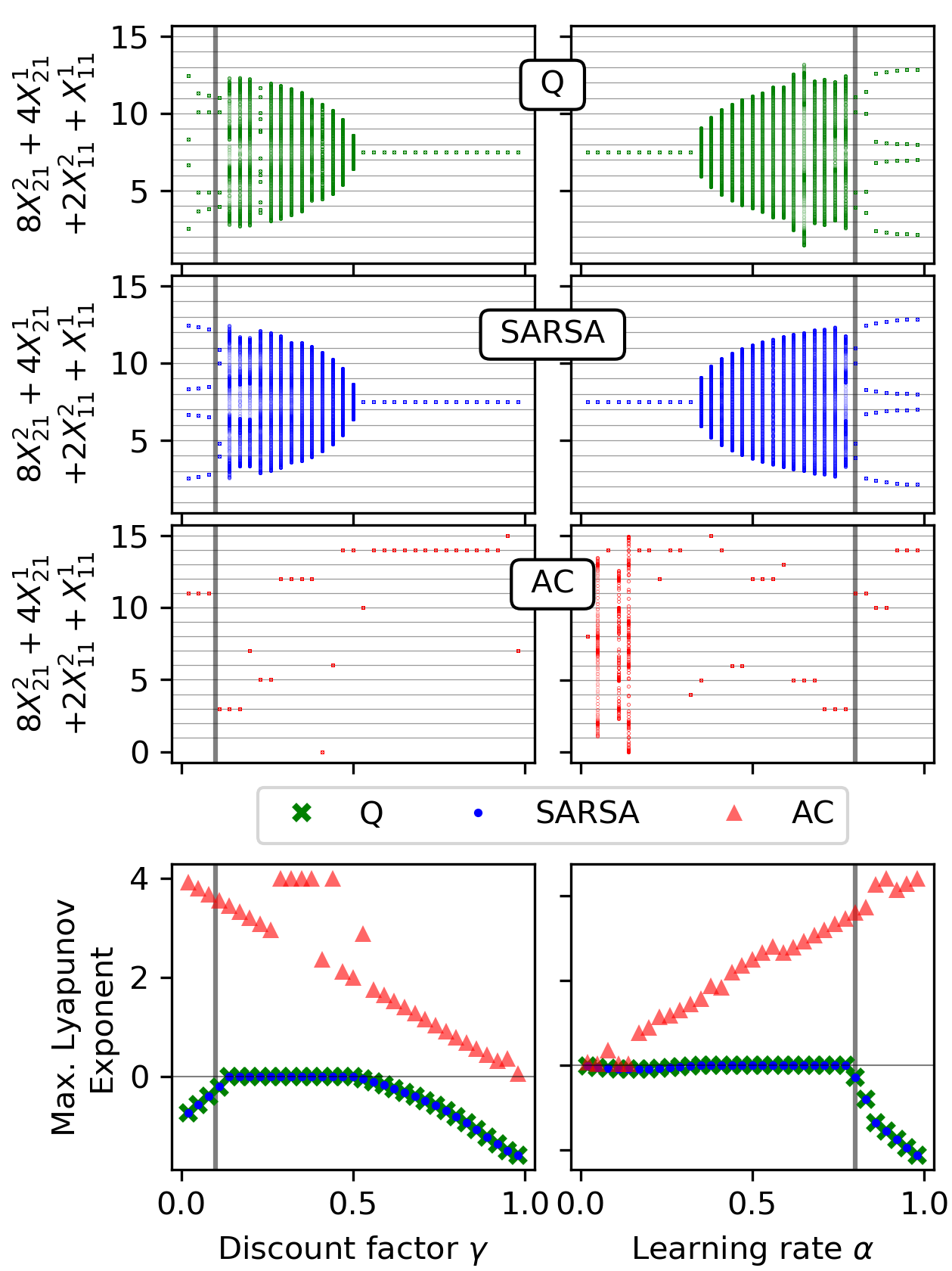}
	\caption{\textbf{Varying discount factor $\gamma$ and learning rate $\alpha$ in  two-state Matching Pennies environment} for intensity of choice $\beta=5.0$ for the Q learners (green crosses), the SARSA learners (blue dots) and the Actor-Critic (AC) learners (red triangles).
	On the left, the discount factor $\gamma$ is varied with learning rate $\alpha=0.8$, as indicated by the gray vertical lines on the right. On the right, the learning rate $\alpha$ is varied with discount factor $\gamma=0.1$ as indicated by the gray vertical lines on the left.
	The three top panels show the visited behavior points during $1000$ iterations after a transient period of $100\, 000$ time steps from initial behavior $(X^1_{11}, X^2_{11}, X^1_{21}, X^2_{21}) = (0.01, 0.99, 0.3, 0.4)$. 
	Visited points are mapped to the function $8X^2_{21}+4X^1_{21}+2X^2_{11}+X^1_{11}$ on the vertical axes to give a fuller image of the visited behavior profiles.
	The bottom panel shows the corresponding largest Lyapunov exponents for the three learners.
	Overall, Q and SARSA learner behave qualitatively more similarly than the Actor-Critic learner.
	}
	\label{fig:MP_BD}
\end{figure}

Specifically, we see that Q learners and SARSA learners behave qualitatively similarly in contrast to the AC learners for both learning rates $\alpha$. 
For the low learning rate $\alpha=0.02$, Q and SARSA learners reach a fixed point of playing both actions with equal probability in both states, yielding a reward of 0.5. Due to the low $\alpha$, this takes approximately 600 time steps.
In contrast, the reward trajectory of the AC learner appears to be chaotic. Figure \ref{fig:MP_BD} confirms this observation, which we will discuss in more detail below.

For the high learning rate $\alpha=0.8$, both Q and SARSA learners enter a periodic limit cycle. Differences in the trajectories of Q and SARSA learner are clearly visible.
The time average reward of this periodic orbit appears to be approximately $0.5$ for each agent, identical to the reward of the fixed point at lower $\alpha$.
The AC learner, however, converges to a fixed point after oscillating near the edges of the phase space. At this fixed point agent 1 plays action 1 in state 1 with probability 1. Thus, it has trapped the system into state 2. In state 2, agent 1 plays action 2 and agent 2 plays action 1 with probability 1 and, consequently, agent 1 receives a reward of 1, whereas agent 2 receives 0 reward. One might ask, Why does agent 2 not decrease their probability for playing action 1, thereby increasing their own reward? And, indeed, the arrows of the temporal difference error suggest this change of behavior profile. 
However, agent 2 cannot follow because their behavior is trapped on the simplex of nonzero action probabilities $X^2_{2a}$. For only $M=2$ actions, $X^2_{21} = 1$ thus can no longer change, regardless of the temporal difference error. 

\begin{figure*}
	\centering
	\includegraphics[width=1.0\linewidth]{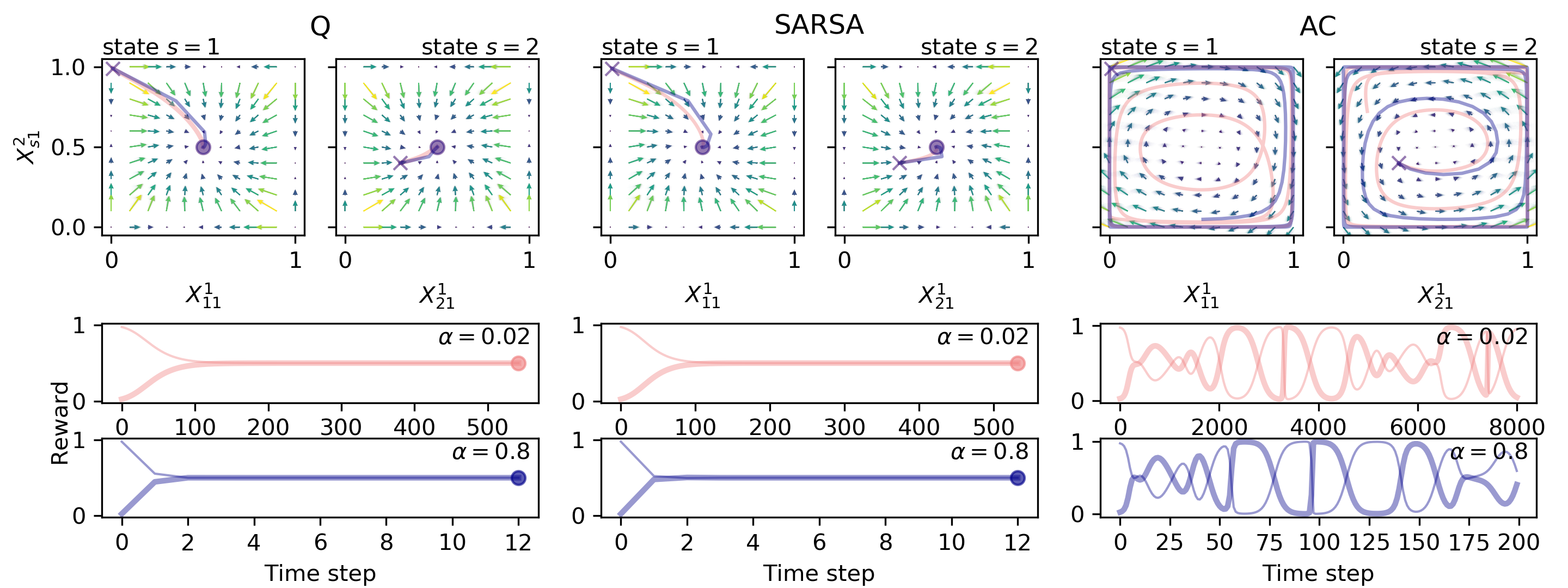}
	\caption{\textbf{Two-state Matching Pennies environment} for high discount factor $\gamma=0.9$; otherwise identical to Fig. \ref{fig:MP_lowgamma}.
	}
	\label{fig:MP_highgamma}
\end{figure*}

Increasing the discount factor to $\gamma = 0.9$, we observe the learning rate $\alpha$ to set the timescale of learning (Fig. \ref{fig:MP_highgamma}). The intensity of choice remained $\beta=5.0$. 
A high learning rate $\alpha=0.8$ corresponds to faster learning in contrast to a low learning rate $\alpha=0.02$. Also, the ratio of learning timescales is comparable to the inverse ratio of learning rates.
For both $\alpha$, Q and SARSA learners reach a fixed point, whereas the AC learners seem to move chaotically (details to be investigated below). 
Comparing the trajectories between the learning rates $\alpha$, we observe a similar shape for each pair of learners. However, the similarity of the AC trajectories decreases at larger time steps.

So far, we varied two parameters: the discount factor $\gamma \in [0, 1)$ and the learning rate $\alpha \in (0, 1)$. Combining Figs. \ref{fig:MP_lowgamma} and \ref{fig:MP_highgamma}, we investigated all four combinations of a low and a high $\gamma$ with a low and a high $\alpha$.
We can summarize that Q and SARSA learners converge to a fixed point for all combinations of discount factor $\gamma$ and learning rate $\alpha$, except when $\gamma$ is low and $\alpha$ simultaneously high. AC dynamics seem chaotic for all combinations of $\alpha$ and $\gamma$.

To investigate the relationship between the parameters more thoroughly, Fig.~\ref{fig:MP_BD} shows bifurcation diagrams with the bifurcation parameters $\alpha$ and $\gamma$. Additionally, it also gives the largest Lyapunov exponents for each learner and each parameter combination. A largest Lyapunov exponent greater than zero is a key characteristic of chaotic motion. We computed the Lyapunov exponent from the analytically derived Jacobian matrix, iteratively used in a QR decomposition according to Ref. \cite{Sandri1996}. See Appendix \ref{sec:Appendix} for details.

The largest Lyapunov exponent for Q and SARSA learners align almost perfectly with each other, whereas the largest Lyapunov exponent of  the AC learners behaves qualitatively different. We first describe the behavior of the Q and SARSA learner: 
For high learning rates $\alpha$ and low farsightedness $\gamma$, Fig.~\ref{fig:MP_BD} shows 
a periodic orbit with few (four) points in phase space. Largest Lyapunov exponents are distinctly below 0 at those regimes. 
Increasing the farsightedness $\gamma$ both learners enter a regime of visiting many points in phase space around the stable fixed point $(X^1_{11}, X^2_{11}, X^1_{21}, X^2_{21}) = (0.5, 0.5, 0.5, 0.5)$. The largest Lyapuonv exponents are close to zero. With increasing $\gamma$ the distance around this fixed-point solution decreases until the dynamics converge from a farsightedness $\gamma$ slightly greater than 0.5 onward. From there the largest Lyapunov exponent decreases again for further increasing $\gamma$. 
The same observations can be made along a decreasing bifurcation parameter $\alpha$, except that at the end, for low $\alpha$, the largest Lyapunov exponents do not decrease as distinctly as for high $\gamma$.

The behavior of the Actor-Critic dynamics is qualitatively different from the one of Q and SARSA. 
The placement of the fixed points on the natural numbers grid suggests that the AC learner get confined on one of the 16 ($M^{NZ}$) corners of the behavior phase space. No regularity to which fixed point the AC learner converges can be deduced. The largest Lyapunov exponent is always above zero and experiences an overall decreasing behavior. 
Similarly, for a decreasing bifurcation parameter $\alpha$, the largest Lyapunov exponent tends to decrease as well. Different from the bifurcation diagram along $\gamma$, for low $\alpha$ the system might enter a periodic motion, but only for some parameters $\alpha$. No regularity can be determined at which parameters $\alpha$ the AC learners enter a periodic motion. 
A more thorough investigation of the nonlinear dynamics, especially those of the Actor-Critic learner, seems of great interest but is, however, beyond the scope of this article and leaves promising paths for future work.  

\begin{figure}
	\centering
	\includegraphics[width=0.98\linewidth]{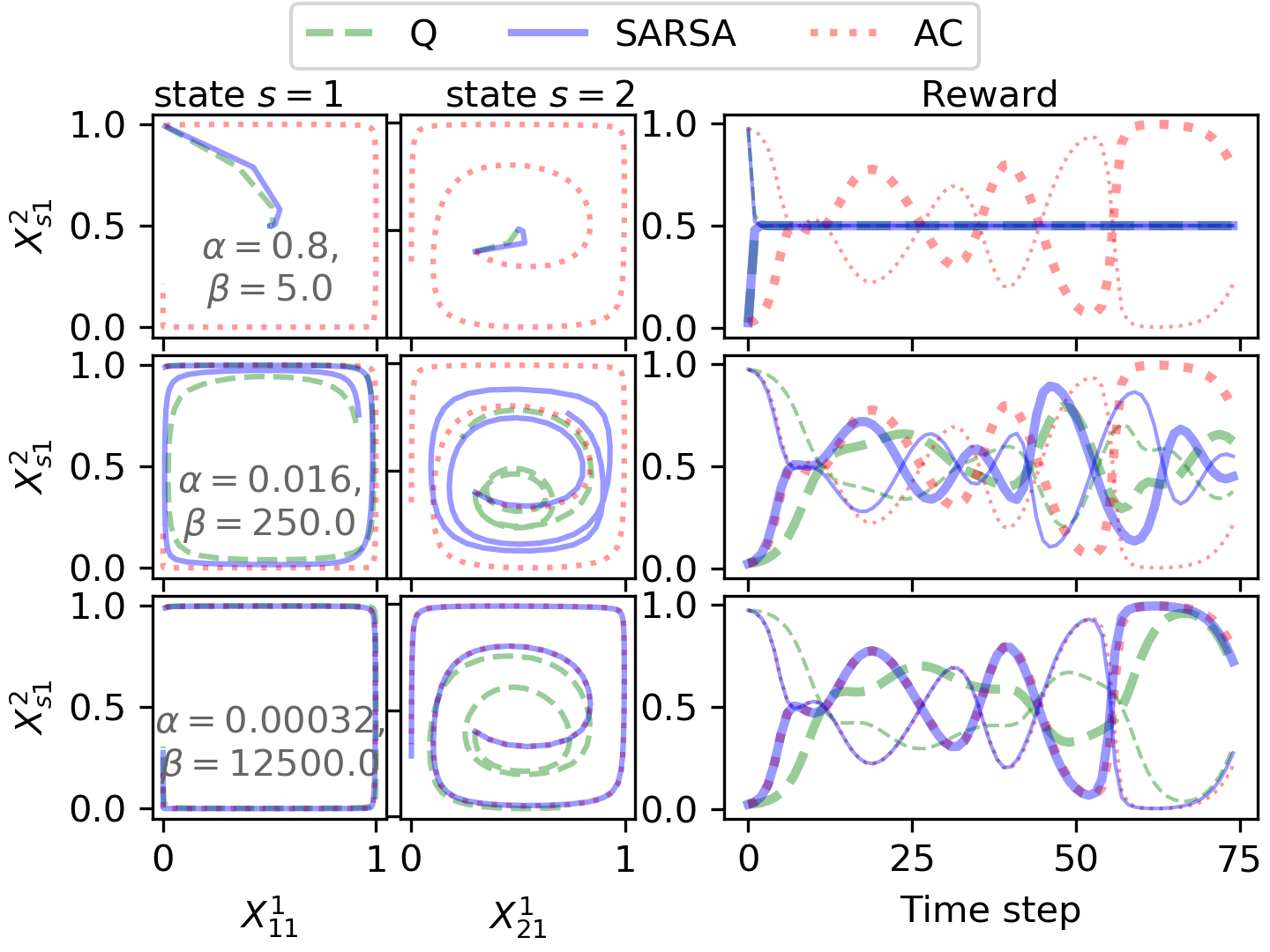}
	\caption{\textbf{Varying intensity of choice $\beta$ under constant $\alpha \cdot \beta$ in two-state Matching Pennies environment} for discount factor $\gamma=0.9$.
	On the left trajectories of the three learners (Q: green dashed, SARSA: blue straight, Actor-Critic(AC): red dotted) are shown in the two behavior space sections, one for each state. On the right, the corresponding reward trajectories are shown.
	The initial behavior was $(X^1_{11}, X^2_{11}, X^1_{21}, X^2_{21}) = (0.01, 0.99, 0.3, 0.4)$.
	The bold reward trajectory belongs to agent 1 and the thin one to agent 2.
	One observes the deterministic limit of Actor-Critic learning to be invariant under constant $\alpha\cdot\beta$ and SARSA learning to converge to AC learning under $\beta\rightarrow\infty$.
	}
	\label{fig:MP_alphabeta}
\end{figure}

Concerning the parameter $\beta$, the intensity of choice, one can infer from the update equations (Eq. \ref{eq:JointPolicyIteration} combined with Eq. \ref{eq:TDe_S} and Eq. \ref{eq:TDe_A}) that the dynamics for the AC learner are invariant for a constant product $\alpha\beta$. 
This is because the temporal difference error of the Actor-Critic learner in the deterministic limit is independent of $\beta$.
Further, the dynamics of the SARSA learner will converge to the dynamics of the AC learner under $\beta \rightarrow \infty$. Figure \ref{fig:MP_alphabeta} nicely confirms these two observations.
Observing Table \ref{tab:LearingEquationsOverview} is another way to see this. Since the value estimate of the future state is identical for SARSA and AC learning, letting the value estimate of the current state vanish by sending $\beta \rightarrow \infty$ makes the SARSA learners approximate the AC learners. 

As mentioned before, $\beta$ controls the exploration-exploitation trade-off. In the temporal difference errors of the Q and SARSA learner it appears in the term indicating the value estimate of the current state $-1/\beta^i \log(X^i_{sa})$. If this term dominates the temporal difference error (i.e., if $\beta$ is small), then the learners tend toward the center of behavior space, i.e., $(X^1_{11}, X^2_{11}, X^1_{21}, X^2_{21}) = (0.5, 0.5, 0.5, 0.5)$, forgetting what they have learned about the obtainable reward.
This characteristic happens to be favorable in our two-state Matching Pennies environment, which is why Q and SARSA learners perform better in finding the $(X^1_{11}, X^2_{11}, X^1_{21}, X^2_{21}) = (0.5, 0.5, 0.5, 0.5)$ solution.
On the other hand, if $\beta$ is large, then the temporal difference error is dominated by the current reward and future value estimate. Not being able to forget, the learners might get trapped in unfavorable behavior, as we can see observing the Actor-Critic learners.
To calibrate $\beta$ it is useful to make oneself clear that it must come in units of [log behavior]\,/\,[reward].

\subsection{Two-state Prisoner's Dilemma}

\begin{figure}[]
	\centering
	\includegraphics[width=0.98\linewidth]{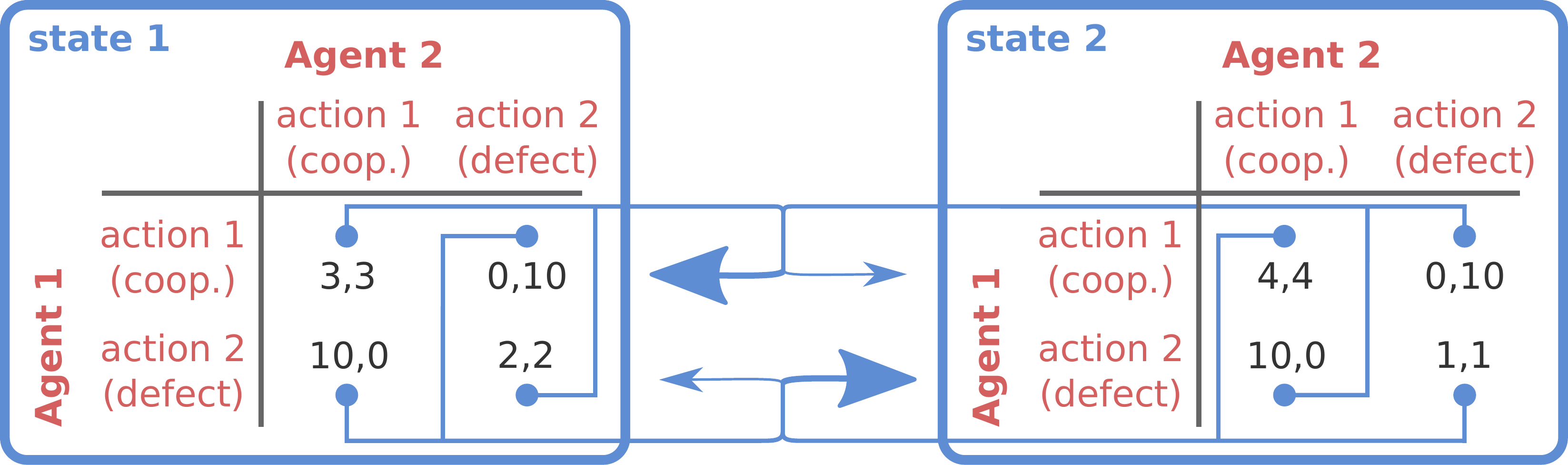}
	\caption{\textbf{Two-state Prisoners Dilemma.} Rewards are given in black type in the payoff tables for each state. State-transition probabilities are indicated by (blue) arrows.	}
	\label{fig:PDgame}
\end{figure}

The single state Prisoners Dilemma is another paradigmatic two-agents, two-actions game. It has been used to model social dilemmas and study the emergence of cooperation.
It describes a situation in which two prisoners are separately interrogated, leaving them with the choice to either cooperate with each other by not speaking to the police or defecting by testifying. 

The two-state version, which has been used as a test environment also in Refs. \cite{VrancxEtAl2008,HennesEtAl2009,HennesEtAl2010}, extends this situation somewhat artificially by playing a Prisoner's Dilemma in each of the two states with a transition probability of 10\% from one state to the other if both agents chose the same action, and a transition probability of 90\% if both agents chose opposite actions.

Figure \ref{fig:PDgame} illustrates these game dynamics. Formally, the payoff matrices are given by
\begin{equation*}
	\left(\begin{smallmatrix}
	R^1_{111s'}, R^2_{111s'} & R^1_{112s'}, R^2_{112s'} \\
	R^1_{121s'}, R^2_{121s'} & R^1_{122s'}, R^2_{122s'}
	\end{smallmatrix}\right) 
	=
	\left(\begin{smallmatrix*}
	3,3 & 0,10 \\
	10,0 & 2,2
	\end{smallmatrix*}\right) 
\end{equation*}
in state $1$
and
\begin{equation*}
	\left(\begin{smallmatrix}
	R^1_{211s'}, R^2_{211s'} & R^1_{212s'}, R^2_{212s'} \\
	R^1_{221s'}, R^2_{221s'} & R^1_{222s'}, R^2_{222s'}
	\end{smallmatrix}\right) 
	=
\left(\begin{smallmatrix}
4,4 & 0,10 \\
10,0 & 1,1
\end{smallmatrix}\right)  
\end{equation*}
in state $2$ for $s' \in \{1, 2\}$, respectively. The corresponding state transition probabilities are given by
\begin{equation*}
	\left(\begin{smallmatrix}
	T_{1112} & T_{1122}\\
	T_{1212} & T_{1222}
	\end{smallmatrix}\right) 
	=
	\left(\begin{smallmatrix}
	T_{2111} & T_{2121}\\
	T_{2211} & T_{2221}
	\end{smallmatrix}\right) 
	= 
	\left(\begin{smallmatrix}
	0.1 & 0.9 \\
	0.9 & 0.1
	\end{smallmatrix}\right).
\end{equation*}

To be precise, the rewards in each state do not resemble a classical social dilemma situation. This is because if both agents would alternately cooperate and defect, both could receive a larger reward per time step compared to always cooperating. Hence, this stochastic game, as it was used in Refs. \cite{VrancxEtAl2008,HennesEtAl2009,HennesEtAl2010}, presents more a coordination than a cooperation challenge to the agents.
The multistate environment can here function as a coordination device.

A behavior profile in which one agent exploits the other in one state, while being exploited in the other state, would result in an average reward per time step of 5 for each agent, e.g., $(X^1_{11}, X^2_{11}, X^1_{21}, X^2_{21}) = (0, 1, 1, 0)$.

\begin{figure*}
	\centering
	\includegraphics[width=1.0\linewidth]{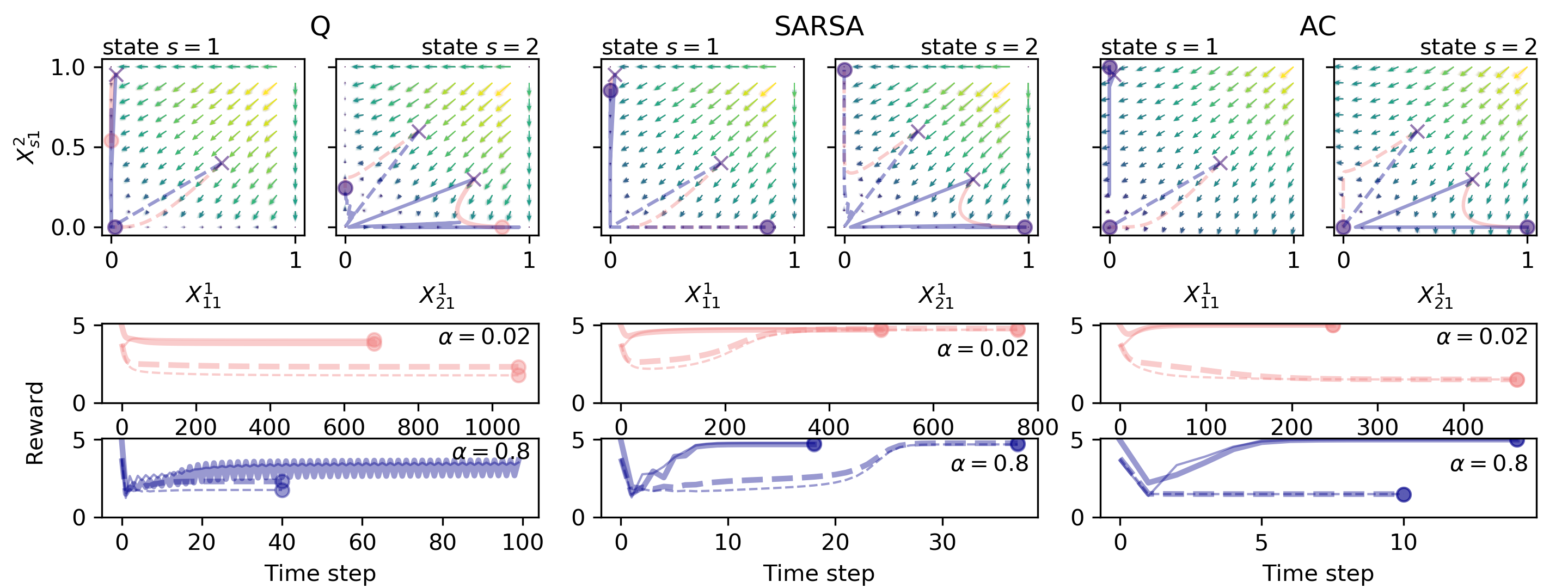}
	\caption{\textbf{Two-state Prisoners Dilemma environment} for discount factor $\gamma=0.45$; otherwise identical to Fig. \ref{fig:MP_lowgamma}.}
	\label{fig:PD_SP_midGamma}
\end{figure*}

However, for all three learning types with a mid-ranged farsightedness ($\gamma = 0.45$) and an intensity of choice $\beta=5.0$, the temporal difference error arrows are pointing on average toward the lower left defection-defection point for each state in behavior phase space (Fig.~\ref{fig:PD_SP_midGamma}). To see whether the three learning types may converge to the described defect-cooperate-cooperate-defect equilibrium, individual trajectories from two exemplary initial conditions and for two learning rates $\alpha$ are shown, a small one ($\alpha=0.02$) and a high one ($\alpha=0.8$).

We observe qualitatively different behavior across all three learners. 
The Q learners converge to equilibria with average rewards distinctly below 5, and the SARSA learners converge to equilibira with average rewards of almost 5 for both learning rates $\alpha$ and both exemplary initial conditions. Both Q and SARSA learners converge to solutions of proper probabilistic behavior, i.e., choosing action cooperate and action defect with nonvanishing chance.  
The Actor-Critic learners, on the other hand, converge to the deterministic defect-cooperate-cooperate-defect behavior described above for the initial condition shown with the non-dashed lines in Fig.~\ref{fig:PD_SP_midGamma} for both learning rates $\alpha$ (shown in light red and dark blue). For the other exemplary initial condition, shown with the dashed lines, it converges to an all-defection solution in both states for both $\alpha$.  

Interestingly, for all learners, all combinations of initial conditions and learning rates converge to a fixed point solution, except for the Q learners with a comparably high learning rate $\alpha=0.8$, which enter a periodic behavior solution for the initial condition with the nondashed line.
The same phenomenon occurred also in the Matching Pennies environment for low farsightedness $\gamma=0.1$, however, there for both Q and SARSA learners. It seems to be caused by the comparably high learning rate. A high learning rate overshoots the behavior update, resulting in a circling behavior around the fixed point.
As in Fig.~\ref{fig:MP_lowgamma}, the time average reward of the periodic orbit seems to be comparable to the reward of the corresponding fixed point at lower $\alpha$.
Furthermore, we observe the same time rescaling effect of the learning rate $\alpha$ in Figure \ref{fig:PD_SP_midGamma} as in Fig.~\ref{fig:MP_highgamma}. 

\begin{figure*}
	\centering
	\includegraphics[width=.9\linewidth]{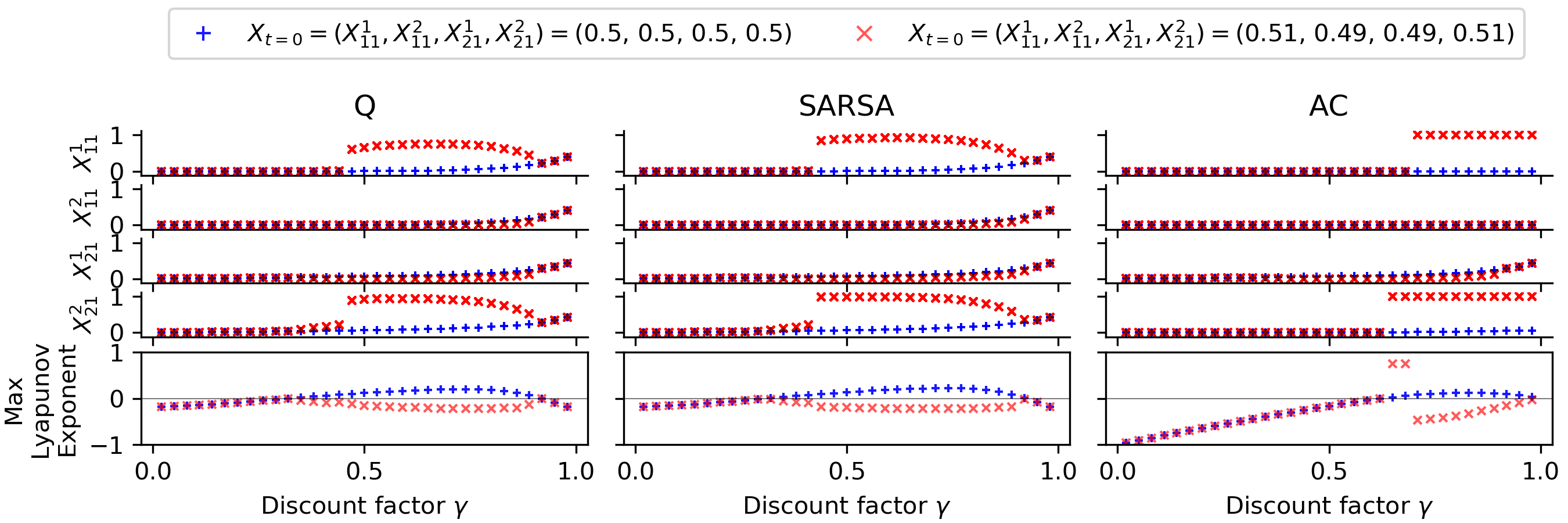}
	\caption{\textbf{Varying discount factor $\gamma$ in  two-state Prisoners Dilemma environment} for learning rate $\alpha=0.2$ and intensity of choice $\beta=5.0$
	 for the Q learners on the left, the SARSA learners in the middle and the Actor-Critic learners on the right.
	The four top panels for each learner show the visited behavior points $X^1_{11}, X^2_{11}, X^1_{21}, X^2_{21}$ during $1000$ iterations after a transient period of $5000$ time steps from initial behavior $(X^1_{11}, X^2_{11}, X^1_{21}, X^2_{21}) = (0.5, 0.5, 0.5, 0.5)$ in blue pluses and from initial behavior $(X^1_{11}, X^2_{11}, X^1_{21}, X^2_{21}) = (0.51, 0.49, 0.49, 0.51)$ in red crosses.
	The bottom panels show the corresponding largest Lyapunov exponents for the two initial conditions.
	Above a critical discount factor $\gamma$ all learners find the high rewarding solution from the red crosses initial condition, but do not do so from the blue pluses initial condition.
	}
	\label{fig:PD_BDalongGamma}
\end{figure*}

To visualize the influence of the discount factor $\gamma$ on the converged behavior, Fig.~\ref{fig:PD_BDalongGamma} shows a bifurcation diagram along the bifurcation parameter $\gamma$ for two initial conditions. Pluses in blue result from a uniformly random behavior profile of $(X^1_{11}, X^2_{11}, X^1_{21}, X^2_{21}) = (0.5, 0.5, 0.5, 0.5)$, whereas the crosses in red initially started from the behavior profile $(X^1_{11}, X^2_{11}, X^1_{21}, X^2_{21}) = (0.51, 0.49, 0.49, 0.51)$.

Across all learners, lower discount factors $\gamma$ correspond to all-defect solutions, whereas for higher $\gamma$ the solutions from the initial condition shown with red crosses tend toward the cooperate-defect-defect-cooperate solution. 
For low $\gamma$, the agents are less aware of the presence of other states and find the all-defect equilibrium solution of the iterated normal form Prisoner's Dilemma. The state transition probabilities have less effect on the learning dynamics. Only above a certain farsightedness do the agents find the more rewarding cooperate-defect-defect-cooperate solution.

The observation from Fig.~\ref{fig:PD_SP_midGamma} is confirmed that the probability to cooperate (i.e., here $X^1_{11}$ and $X^2_{21}$ ) is lowest for the Q learners, midrange for the SARSA learners and 1 for the Actor-Critic learners. 
One reason for this observation can be found in the intensity of choice parameter $\beta$. It balances the reward obtainable in the current behavior space segment with the forgetting of current knowledge to be open to new solutions. Such forgetting expresses itself by temporal difference error components pointing toward the center of behavior space. Thus, a relatively small $\beta=5.0$ can explain why solutions at the edge of the behavior space cannot be reached by Q and SARSA learners. The AC learner misses this forgetting term in the deterministic limit and can therefore easily enter behavior profiles at the edge of the behavior space. 

Q and SARSA learners have a critical discount factor $\gamma$ above which the cooperate-defect-defect-cooperate high reward solution is obtained and below which the all-defect low reward solution gets selected. 
However, for increasing discount factors $\gamma$ up to 1, Q and SARSA learners experience a drop in playing the cooperative action probability.

The Actor-Critic learners approach the cooperate-defect-defect-cooperate solution in two steps. For increasing $\gamma$, first the probability of agent 2 cooperating in state 2 ($X^2_{21}$) jumps from zero to 1 while agent 1 still defects in state 1. Only after a slight increase of $\gamma$ does agent 1 also cooperate in state 1 ($X^1_{11}$). 

Interestingly, for the uniformly random initial behavior condition shown with blue pluses, there is no critical discount factor $\gamma$ and no learners come close to the cooperate-defect-defect-cooperate solution. Here, only for $\gamma$ close to 1 do all cooperation probabilities $X^{i}_{s1}$ gradually increase.
Furthermore, exactly at those $\gamma$, where the cooperate-defect-defect-cooperate solution is obtained from the initial behavior condition shown with red crosses, the solutions from the uniformly random initial behavior condition (blue pluses) have a largest Lyapuonv exponent greater than 0. At other values of $\gamma$, the largest Lyapunov exponents for the two initial conditions overlap. This suggests that the largest Lyapunov exponents greater than zero may point to the fact that other, perhaps more rewarding solutions may exist in phase space. 
A more thorough investigation regarding this multistability is an open point for future research.

\begin{figure}
	\centering

	\includegraphics[width=.9\linewidth]{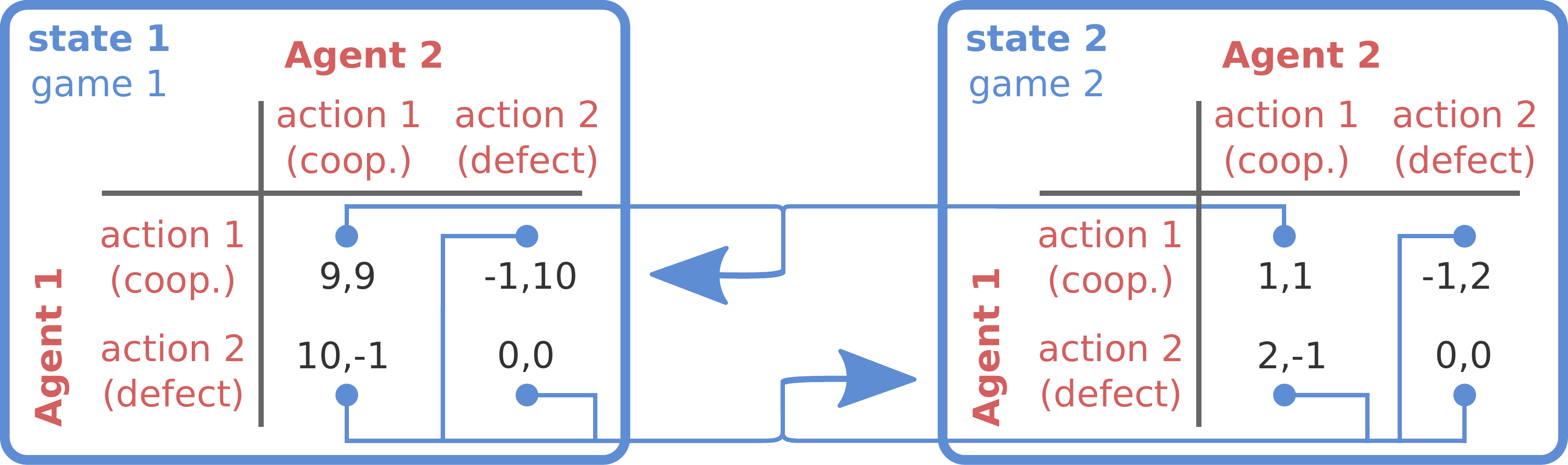}
		
		\vspace{0.1cm}
		
	\includegraphics[width=.99\linewidth]{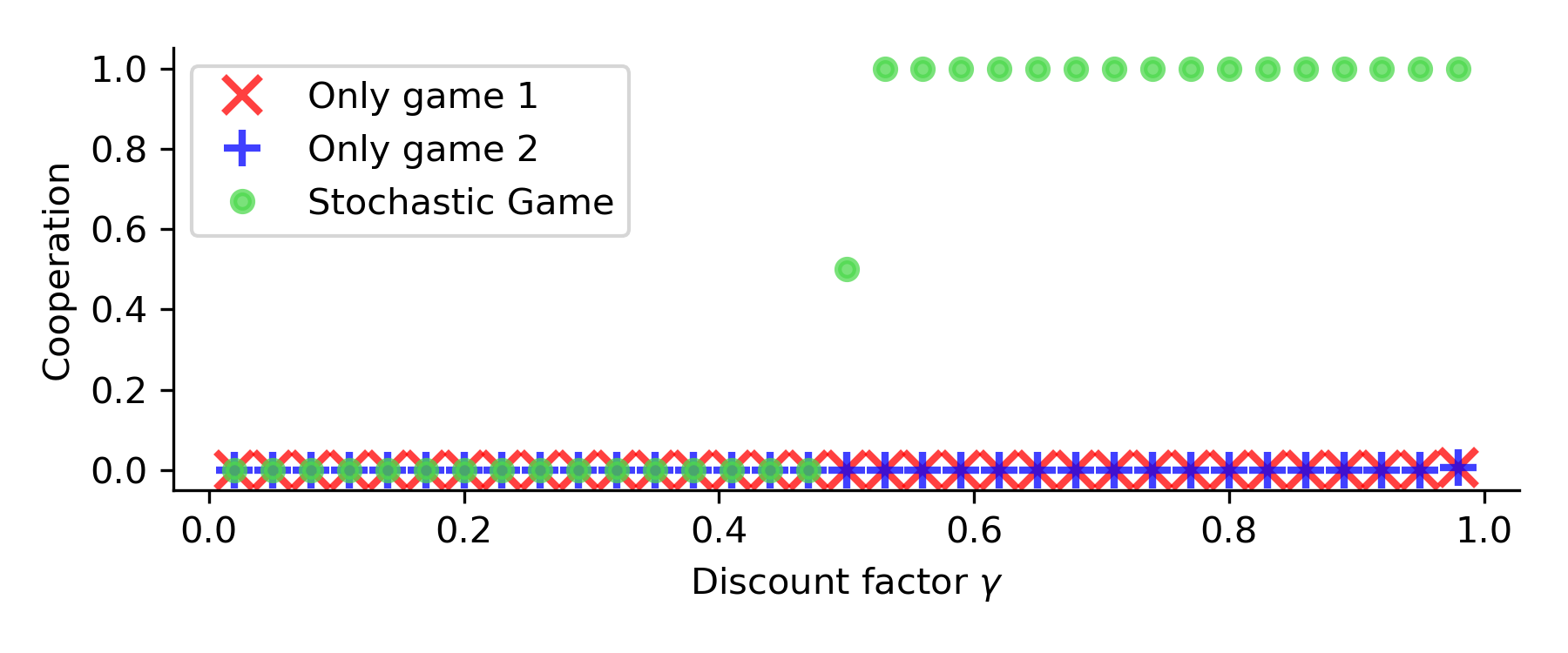}
	
	\caption{\textbf{Cooperation challenge in a two-state Prisoner's Dilemma}.
		Top panel shows a two-state Prisoner's Dilemma game, whose state games individually favor defection. Bottom panels shows the level of cooperation SARSA learners with $\alpha=0.016$, $\beta=250$ play after reaching a fixed point from the center of behavior space ($X^i_{sa} = 0.5$ for all $i,s,a$) for varying discount factors $\gamma$.
		Results for Q and AC learners are similar. Cooperation levels are shown for the full stochastic game as well as for each individual state game played repeatedly. 
		For sufficiently large farsightedness, cooperation can emerge in the stochastic game, in contrast to the individual repeated games. 
	}
	\label{fig:PD_v2}
\end{figure}

As we have argued above, the two-state Prisoner's Dilemma as it was used in Refs. \cite{VrancxEtAl2008,HennesEtAl2009,HennesEtAl2010} presents rather a coordination than a cooperation challenge to the agents.
Figure \ref{fig:PD_v2} demonstrates that our learning dynamics are also capable of solving a cooperation challenge in a stochastic game setting, for which we adapt a two-state Prisoner's Dilemma in analogy to Ref. \cite{HilbeEtAl2018}.
Figure \ref{fig:PD_v2} confirms previous findings that cooperation emerges only in the stochastic game, compared to playing each Prisoner's Dilemma repeatedly \cite{HilbeEtAl2018}. Further, cooperation only emerges for sufficiently large farsightedness $\gamma$.

\section{Discussion}
\label{sec:Discussion}

The main contribution of this paper is the development of a technique to obtain the deterministic limit of temporal difference reinforcement learning.
Through our work we have combined the literature on learning dynamics from statistical physics with the evolutionary game theory-inspired learning dynamics literature from machine learning.
For the statistical physics community, the novelty consists of learning equations, capable of handling environmental state transitions. For the machine learning community the novelty lies in the systematic methodology we have used to obtain the deterministic learning equations.

We have demonstrated our approach with the three prominent reinforcement learning algorithms from computer science: Q learning, SARSA learning, and Actor-Critic learning.
A comparison of their dynamics in previously used two-agent, two-actions, two-states environments has revealed the existence of a variety of qualitatively different dynamical regimes, such as convergence to fixed points, periodic orbits and deterministic chaos.

We have found that
Q and SARSA learners tend to behave qualitatively more similar in comparison to the Actor-Critic learning dynamics. 
This characteristic results at least partly from our relatively low intensity of choice parameter $\beta$, controlling the exploration-exploitation trade-off via a forgetting term in the temporal difference errors.
Sending $\beta \rightarrow \infty$, the SARSA learning dynamics approach the Actor-Critic learning dynamics, as we have shown.
Overall the Actor-Critic learners have a tendency to enter confining behavior profiles, due to their nonexisting forgetting term. This characteristic leaves them trapped at the edges of the behavior space. In contrast, Q and SARSA learner do not show such learning behavior. 
Interestingly, this characteristic of the AC learners turns out to be favorable in the two-state Prisoner's Dilemma environment, where they find the most rewarding solution in more cases compared to Q and SARSA but hinders the convergence to the fixed point solution in the two-state Matching Pennies environment. 
	Thus, the most favorable level of forgetting depends on the environment.
	In order to tune the respective parameter $\beta$, our consideration that it must come in the unit of [log behavior]\,/\,[reward] may be helpful.

We have demonstrated the effect of the learning rate $\alpha$ adjusting the speed of learning by controlling the amount of new information used in a behavior profile update. Thereby, within limits, $\alpha$ functions as a time rescaling.
However, a comparably large learning rate $\alpha$ might cause an overshooting phenomenon, hindering the convergence to a fixed point. Instead, the learners enter a limit cycle around that point. Nevertheless, the average reward of the limit-cycling behavior was approximately equal to the one of the fixed point obtained at lower $\alpha$ but took fewer time steps to reach. 
Thus, perhaps other dynamical regimes than fixed points, such as limit cycles or strange attractors, could be of interest in some applications of reinforcement learning. 

We have also shown the effect of the discount factor $\gamma$ adjusting the farsightedness of the agents. At low $\gamma$ the state transition probabilities have less effect on the learning dynamics compared to high discount factors.  

To summarize the three parameters $\alpha, \beta$, and $\gamma$: The level of exploitation $\beta$ and the farsightedness $\gamma$ control \textit{where} the learner adapts toward in behavior space, weighting current reward, expected future reward and the level of forgetting. The learning rate $\alpha$ controls \textit{how fast} the learner adapts along these directions.

We hope that our work might turn out useful for the application of reinforcement learning in various domains, with respect to 
parameter tuning, 
the design of new algorithms, and
the analysis of complex strategic interactions using meta strategies, as Bloembergen\textit{ et al.} \cite{BloembergenEtAl2015} have pointed out. 
In this regard, future work could extend the presented methodology to
partial observability of the Markov states of the environment \cite{Spaan2012,Oliehoek2012},
behavior profiles with history,
and other-regarding agent (i.e. joint-action) learners (cf. Ref. \cite{BusoniuEtAl2008} for an overview of other-regarding agent learning algorithms).
Also, the combination of individual reinforcement learning and social learning through imitation \cite{Bandura1977,SmollaEtAl2015,BarfussEtAl2017,BanischOlbrich2018} seems promising. Such endeavors would naturally lead to the exploration of network effects. It is important to note that only a few dynamical systems reinforcement learning studies have begun to incorporate network structures between agents \cite{BladonGalla2011,RealpeGomezEtAl2012}.

Apart from these more technical extensions, we hope that our learning equations will prove themselves useful when studying the evolution of cooperation in stochastic games \cite{HilbeEtAl2018}. With stochastic games one is able to explicitly account for a changing environment.
Therefore, such studies are likely to contribute to the advancement of theoretical research on the sustainability of interlinked social-ecological systems \cite{Levin2013,DongesEtAl2017}.
Interactions, synergies, and trade-offs between social \cite{Dawes1980,MacyFlache2002} and ecological \cite{HeitzigEtAl2016} dilemmas can be explored using the framework of stochastic games.
More realistic environments, modeling e.g., the harvesting of common-pool renewable resources \cite{LindkvistNorberg2014,SchillEtAl2015} or the prevention of dangerous climate change \cite{MilinskiEtAl2008,BarrettDannenberg2012}, for our learning dynamics are likely to prove themselves useful.
Here, it may be of interest to evaluate the learning process not only in terms of efficiency but also how close it came to the optimal behavior.
Other paradigms than value optimization may also be important \cite{BarfussEtAl2018}, such as sustainability or resilience \cite{DongesBarfuss2017}.

\footnotesize{
Python code to reproduce the figures of this article is available at
\url{https://doi.org/10.5281/zenodo.1495091}. }

\begin{acknowledgments}
This work was developed in the context of the COPAN project on Coevolutionary
Pathways in the Earth System at the Potsdam Institute for Climate
Impact Research (PIK). We are grateful for financial support by the
Heinrich B\"oll Foundation, the
Stordalen Foundation (via the Planetary Boundaries Research Network PB.net), the
Earth League's EarthDoc program and the Leibniz Association (project DOMINOES).
We thank Jobst Heitzig for discussions and comments on the manuscript.
\end{acknowledgments}

\appendix

\begin{widetext}
\section{Computation of Lyapunov Exponents}
\label{sec:Appendix}
We compute the Lyapunov exponents using an iterative QR decomposition of the Jacobian matrix according to Sandri \cite{Sandri1996}. In the following we present the derivation of the Jacobian matrix.

Eq. \ref{eq:JointPolicyIteration} constitutes a map $f$, which iteratively updates the behavior profile $\textbf{X} \in \mathds R^{N\times M \times Z}$. Consequently, we can represent its derivative as a Jacobian tensor $f'(\textbf{X}) \in \mathds R^{N\times M \times Z \times N\times M \times Z}$.

Let $A^i_{sa}:= X^i_{sa}  \exp\Big(\alpha^i\beta^i T\!D^i_{sa}(\mathbf X)\Big)$ be the numerator of Eq. \ref{eq:JointPolicyIteration}, and $B^i_s:= \sum_b A^i_{sb}$ its denominator, i.e. $f =: A/B$. Hence, 
\begin{equation}
f'(\mathbf X) = \frac{A'B - B'A}{B^2}
\end{equation}
or, more precisely, in components, 
\begin{equation}
\frac{df^i_{sa}(\mathbf X)}{dX^j_{rb}} =
\frac{\frac{dA^i_{sa}(\mathbf X)}{dX^j_{rb}} B^i_s(\mathbf X) 
- \frac{dB^i_s}{dX^j_{rb}}(\mathbf X) A^i_{sa}(\mathbf X)}{(B^i_s(\mathbf X))^2}.
\end{equation}

$A$ and $B$ are known, and if $A'$ is known, then $B'$ is easily obtained by $ \frac{dB^i_{s}(\mathbf X)}{dX^j_{rb}} = \sum_c \frac{dA^i_{sc}(\mathbf X)}{dX^j_{rb}}$. Therefore we need to compute $A'$ for the three learner types Q, SARSA and Actor-Critic learning.

\subsection{Q learning}
Let us rewrite $A^i_{sa}$ for the Q learner according to
\begin{equation}
A^i_{sa}:= (X^i_{sa})^{(1-\alpha^i)}  \exp\Big(\alpha^i\beta^i \hat{T\!D}^i_{sa}(\textbf{X})\Big),
\label{eq:A_Q}
\end{equation}
where we removed the estimate of the current value from the temporal difference error, leaving the truncated TD error as
\begin{equation}
\hat{T\!D}^i_{sa}(\textbf{X}) := (1-\gamma^i) \AvgSpAmI{R}^i_{sa} + \gamma^i
\prescript{\text{max}}{}{\mathcal Q}^i_{sa}(\mathbf X).
\label{eq:truncedTD_Q}
\end{equation}
Hence, we can write the derivative of $A$ as

\begin{equation}
\frac{dA^i_{sa}(\mathbf X)}{dX^j_{rb}} = \exp\Big(\alpha^i\beta^i \hat{T\!D}^i_{sa}(\textbf{X})\Big)
\left((1-\alpha^i)(X^i_{sa})^{-\alpha^i}\frac{dX^i_{sa}}{dX^j_{rb}} + \alpha^i\beta^i (X^i_{sa})^{(1-\alpha^i)} \frac{d\hat{T\!D}^i_{sa}(\textbf{X})}{dX^j_{rb}} \right).
\end{equation}

Since $\sum_c X^i_{sc} = 1$, $dX^i_{sa} / dX^j_{rb}$ can be expressed as
\begin{equation}
\frac{dX^i_{sa} }{ dX^j_{rb} }= \delta_{ij}\delta_{sr} (2\delta_{ab} - 1).
\label{eq:dXdX}
\end{equation}

The derivative of the truncated temporal difference error reads
\begin{equation}
\frac{d\hat{T\!D}^i_{sa}(\textbf{X})}{dX^j_{rb}} =
(1-\gamma^i) \frac{d\AvgSpAmI{R}^i_{sa}}{dX^j_{rb}} +
\gamma^i \frac{d \prescript{\text{max}}{}{\mathcal Q}^i_{sa}(\mathbf X)}{dX^j_{rb}}.
\end{equation}

Let us write the derivative of the reward as
\begin{equation}
\frac{d\AvgSpAmI{R}^i_{sa}}{dX^j_{rb}} = \sum_{s'} \sum_{\mathbf a^{-i}} 
\frac{d \mathbf X^{-i}_{s\mathbf a^{-i}} }{dX^j_{rb}} 
T_{sa\mathbf a^{-i} s'} R^i_{sa\mathbf a^{-i} s'}
\end{equation}
using Eq. \ref{eq:AvgAmi} and Eq. \ref{eq:AvgSp}, where the derivatives $d \mathbf X^{-i}_{s\mathbf a^{-i}} /dX^j_{rb}$ need to be executed according to Eq. \ref{eq:dXdX}.

For the derivative of the maximum next value we write accordingly

\begin{equation}
\frac{d \prescript{\text{max}}{}{\mathcal Q}^i_{sa}(\mathbf X)}{dX^j_{rb}}
= \sum_{s'} \sum_{\mathbf a^{-i}}  \frac{d \mathbf X^{-i}_{s\mathbf a^{-i}}}{dX^j_{rb}} T_{sa \mathbf a^{-i} s'} \max_c Q^i_{s'c}(\mathbf X)
+  \sum_{s'} \sum_{\mathbf a^{-i}}  \mathbf X^{-i}_{s\mathbf a^{-i}} T_{sa \mathbf a^{-i} s'} \frac{d \max_c Q^i_{s'c}(\mathbf X)}{dX^j_{rb}}.
\end{equation}

Let $a^m := \arg\max_a Q^i_{sa}(\mathbf X)$, then
\begin{equation}
\frac{d \max_c Q^i_{sc}(\mathbf X)}{dX^j_{rb}} = \delta_{aa^m} \frac{d Q^i_{sa}(\mathbf X)}{dX^j_{rb}}
\end{equation}

and 
\begin{equation}
\frac{d Q^i_{sa}(\mathbf X)}{dX^j_{rb}} = 
(1-\gamma^i) \frac{d\AvgSpAmI{R}^i_{sa}}{dX^j_{rb}} +
\gamma^i \sum_{s'}
\frac{d \AvgA{T}_{ss'}}{dX^j_{rb}} V^i_{s'}(\textbf{X})  +
\AvgA{T}_{ss'} \frac{dV^i_{s'}(\textbf{X})}{dX^j_{rb}}.
\end{equation}

For the derivative of the effective Markov Chain transition tensor we can write
\begin{equation}
\frac{d \AvgA{T}_{ss'}}{dX^j_{rb}} = \sum_{\mathbf a} 
\frac{d \mathbf X_{s\mathbf a} }{dX^j_{rb}} 
T_{sa\mathbf a^{-i} s'} ,
\end{equation}
using Eqs. \ref{eq:AvgAmi} and \ref{eq:AvgSp}, where again the derivatives $d \mathbf X_{s\mathbf a} /dX^j_{rb}$ need to be executed according to Eq. \ref{eq:dXdX}. 

For the derivative of the state value let us rewrite Eq. \ref{eq:StateValueComputation} as $V^i_s = (1-\gamma^i) \sum_{s'} M^{-1}_{ss'} \AvgSpA{R}^i_{s'}$ with $M := (\mathds 1_Z - \gamma^i \AvgA{T})$. Thus,

\begin{equation}
\frac{dV^i_{s}(\textbf{X})}{dX^j_{rb}}
= (1-\gamma^i) \sum_{s''}
\frac{d (M^{-1}_{ss''})}{dX^j_{rb}} \AvgSpA{R}^i_{s''}
+ M^{-1}_{ss''} \frac{d \AvgSpA{R}^i_{s''} }{dX^j_{rb}}.
\label{eq:dV}
\end{equation}

To obtain the derivative of the inverse matrix $M^{-1}$ we use $(M^{-1} M)' = 0 = (M^{-1})' M + M^{-1} M'$ and therefore $(M^{-1})' = -M^{-1} M' M^{-1}$. For $M'$ we write,
\begin{equation}
\frac{dM_{ss'}}{dX^j_{rb}} = -\gamma^i \frac{d \AvgA{T}_{ss'}}{dX^j_{rb}}.
\end{equation}

We obtain the derivative of the reward according to
\begin{equation}
\frac{d\AvgSpA{R}^i_{s}}{dX^j_{rb}} =
\sum_{s'} \sum_{\mathbf a} 
\frac{d \mathbf X_{s\mathbf a} }{dX^j_{rb}} 
T_{s\mathbf a s'} R^i_{s\mathbf a s'},
\end{equation}
using Eq. \ref{eq:AvgA} and Eq. \ref{eq:AvgSp}, where the derivatives $d X^{i}_{sa}  /dX^j_{rb}$ need to be executed according to Eq. \ref{eq:dXdX}.

Now we can compute the Jacobian matrix for the Q learning dynamics in their deterministic limit.

\subsection{SARSA learning}
The computation of the Jacobian matrix for the SARSA learning update in its deterministic limit is similar, except the truncated temporal difference error reads
\begin{equation}
\hat{T\!D}^i_{sa}(\textbf{X}) := (1-\gamma^i) \AvgSpAmI{R}^i_{sa} + \gamma^i
\prescript{\text{next}}{}{\mathcal Q}^i_{sa}(\mathbf X).
\end{equation}
instead of Eq. \ref{eq:truncedTD_Q}. Hence,
\begin{equation}
\frac{d\hat{T\!D}^i_{sa}(\textbf{X})}{dX^j_{rb}} =
(1-\gamma^i) \frac{d\AvgSpAmI{R}^i_{sa}}{dX^j_{rb}} +
\gamma^i \frac{d \prescript{\text{next}}{}{\mathcal Q}^i_{sa}(\mathbf X)}{dX^j_{rb}},
\end{equation}
and
\begin{equation}
\frac{d \prescript{\text{next}}{}{\mathcal Q}^i_{sa}(\mathbf X)}{dX^j_{rb}}
= \sum_{s'} \sum_{\mathbf a^{-i}}  \frac{d \mathbf X^{-i}_{s\mathbf a^{-i}}}{dX^j_{rb}} T_{sa \mathbf a^{-i} s'} \sum_c X^i_{s'c} Q^i_{s'c}(\mathbf X)
+  \sum_{s'} \sum_{\mathbf a^{-i}}  \mathbf X^{-i}_{s\mathbf a^{-i}} T_{sa \mathbf a^{-i} s'} \frac{d [\sum_c X^i_{s'c} Q^i_{s'c}(\mathbf X)]}{dX^j_{rb}}.
\end{equation}
The derivative of $\sum_c X^i_{s'c} Q^i_{s'c}(\mathbf X)$ reads
\begin{equation}
\frac{d [\sum_c X^i_{s'c} Q^i_{s'c}(\mathbf X)]}{dX^j_{rb}}
=
\sum_c
\left(
\frac{dX^i_{s'c}}{dX^j_{rb}} Q^i_{s'c}(\mathbf X)
+ X^i_{s'c}\frac{dQ^i_{s'c}}{dX^j_{rb}}
\right).
\end{equation}
All remaining terms have already been given in the previous section for the Q learner Jacobian matrix.

\subsection{Actor-Critic learning}
For the Actor-Critic learning update, Eq. \ref{eq:A_Q} reads
\begin{equation}
A^i_{sa}:= X^i_{sa}  \exp\Big(\alpha^i\beta^i \hat{T\!D}^i_{sa}(\textbf{X})\Big),
\label{eq:A_A}
\end{equation}
with the truncated temporal difference error
\begin{equation}
\hat{T\!D}^i_{sa}(\textbf{X}) := (1-\gamma^i) \AvgSpAmI{R}^i_{sa} + \gamma^i
\prescript{\text{next}}{}{\mathcal V}^i_{sa}(\mathbf X).
\end{equation}

The derivative of the next value estimate is obtained by
\begin{equation}
\frac{d \prescript{\text{next}}{}{\mathcal V}^i_{sa}(\mathbf X)}{dX^j_{rb}}
= \sum_{s'} \sum_{\mathbf a^{-i}}  \frac{d \mathbf X^{-i}_{s\mathbf a^{-i}}}{dX^j_{rb}} T_{sa \mathbf a^{-i} s'} V^i_{s'}(\mathbf X)
+  \sum_{s'} \sum_{\mathbf a^{-i}}  \mathbf X^{-i}_{s\mathbf a^{-i}} T_{sa \mathbf a^{-i} s'} \frac{d V^i_{s'}(\mathbf X)}{dX^j_{rb}}.
\end{equation}
The derivative of the next value $V^i_{s'}$ is given by Eq. \ref{eq:dV}.
These are all terms necessary to compute the Jacobian matrix for the Actor-Critic learning update.

\end{widetext}

\input{BarfussEtAlInPrep_DetRL_a.bbl}

\end{document}

%% file: BarfussEtAlInPrep_DetRL_a.bbl
%

%% file: BarfussEtAlInPrep_DetRL_a.bbl
\begin{thebibliography}{60}%
\makeatletter
\providecommand \@ifxundefined [1]{%
 \@ifx{#1\undefined}
}%
\providecommand \@ifnum [1]{%
 \ifnum #1\expandafter \@firstoftwo
 \else \expandafter \@secondoftwo
 \fi
}%
\providecommand \@ifx [1]{%
 \ifx #1\expandafter \@firstoftwo
 \else \expandafter \@secondoftwo
 \fi
}%
\providecommand \natexlab [1]{#1}%
\providecommand \enquote  [1]{``#1''}%
\providecommand \bibnamefont  [1]{#1}%
\providecommand \bibfnamefont [1]{#1}%
\providecommand \citenamefont [1]{#1}%
\providecommand \href@noop [0]{\@secondoftwo}%
\providecommand \href [0]{\begingroup \@sanitize@url \@href}%
\providecommand \@href[1]{\@@startlink{#1}\@@href}%
\providecommand \@@href[1]{\endgroup#1\@@endlink}%
\providecommand \@sanitize@url [0]{\catcode `\\12\catcode `\$12\catcode
  `\&12\catcode `\#12\catcode `\^12\catcode `\_12\catcode `\%12\relax}%
\providecommand \@@startlink[1]{}%
\providecommand \@@endlink[0]{}%
\providecommand \url  [0]{\begingroup\@sanitize@url \@url }%
\providecommand \@url [1]{\endgroup\@href {#1}{\urlprefix }}%
\providecommand \urlprefix  [0]{URL }%
\providecommand \Eprint [0]{\href }%
\providecommand \doibase [0]{http://dx.doi.org/}%
\providecommand \selectlanguage [0]{\@gobble}%
\providecommand \bibinfo  [0]{\@secondoftwo}%
\providecommand \bibfield  [0]{\@secondoftwo}%
\providecommand \translation [1]{[#1]}%
\providecommand \BibitemOpen [0]{}%
\providecommand \bibitemStop [0]{}%
\providecommand \bibitemNoStop [0]{.\EOS\space}%
\providecommand \EOS [0]{\spacefactor3000\relax}%
\providecommand \BibitemShut  [1]{\csname bibitem#1\endcsname}%
\let\auto@bib@innerbib\@empty
\bibitem [{\citenamefont {Sutton}\ and\ \citenamefont
  {Barto}(1998)}]{SuttonBarto1998}%
  \BibitemOpen
  \bibfield  {author} {\bibinfo {author} {\bibfnamefont {Richard~S}\
  \bibnamefont {Sutton}}\ and\ \bibinfo {author} {\bibfnamefont {Andrew~G}\
  \bibnamefont {Barto}},\ }\href@noop {} {\emph {\bibinfo {title}
  {Reinforcement learning: An introduction}}}\ (\bibinfo  {publisher} {MIT
  Press},\ \bibinfo {year} {1998})\BibitemShut {NoStop}%
\bibitem [{\citenamefont {Busoniu}\ \emph {et~al.}(2008)\citenamefont
  {Busoniu}, \citenamefont {Babuska},\ and\ \citenamefont
  {Schutter}}]{BusoniuEtAl2008}%
  \BibitemOpen
  \bibfield  {author} {\bibinfo {author} {\bibfnamefont {L.}~\bibnamefont
  {Busoniu}}, \bibinfo {author} {\bibfnamefont {R.}~\bibnamefont {Babuska}}, \
  and\ \bibinfo {author} {\bibfnamefont {B.~De}\ \bibnamefont {Schutter}},\
  }\bibfield  {title} {\enquote {\bibinfo {title} {A comprehensive survey of
  multiagent reinforcement learning},}\ }\href {\doibase
  10.1109/tsmcc.2007.913919} {\bibfield  {journal} {\bibinfo  {journal} {{IEEE}
  Transactions on Systems, Man, and Cybernetics, Part C (Applications and
  Reviews)}\ }\textbf {\bibinfo {volume} {38}},\ \bibinfo {pages} {156--172}
  (\bibinfo {year} {2008})}\BibitemShut {NoStop}%
\bibitem [{\citenamefont {Wiering}\ and\ \citenamefont {van
  Otterlo}(2012)}]{WieringVanOtterlo2012}%
  \BibitemOpen
  \bibfield  {author} {\bibinfo {author} {\bibfnamefont {M}~\bibnamefont
  {Wiering}}\ and\ \bibinfo {author} {\bibfnamefont {Merendel}\ \bibnamefont
  {van Otterlo}},\ }\href {\doibase 10.1007/978-3-642-27645-3} {\emph {\bibinfo
  {title} {Reinforcement Learning: State-of-the-Art}}}\ (\bibinfo  {publisher}
  {Springer Berlin Heidelberg},\ \bibinfo {year} {2012})\BibitemShut {NoStop}%
\bibitem [{\citenamefont {Shah}(2012)}]{Shah2012}%
  \BibitemOpen
  \bibfield  {author} {\bibinfo {author} {\bibfnamefont {Ashvin}\ \bibnamefont
  {Shah}},\ }\bibfield  {title} {\enquote {\bibinfo {title} {Psychological and
  neuroscientific connections with reinforcement learning},}\ }in\ \href
  {\doibase 10.1007/978-3-642-27645-3_16} {\emph {\bibinfo {booktitle}
  {Reinforcement Learning}}}\ (\bibinfo  {publisher} {Springer Berlin
  Heidelberg},\ \bibinfo {year} {2012})\ pp.\ \bibinfo {pages}
  {507--537}\BibitemShut {NoStop}%
\bibitem [{\citenamefont {Hassabis}\ \emph {et~al.}(2017)\citenamefont
  {Hassabis}, \citenamefont {Kumaran}, \citenamefont {Summerfield},\ and\
  \citenamefont {Botvinick}}]{HassabisEtAl2017}%
  \BibitemOpen
  \bibfield  {author} {\bibinfo {author} {\bibfnamefont {Demis}\ \bibnamefont
  {Hassabis}}, \bibinfo {author} {\bibfnamefont {Dharshan}\ \bibnamefont
  {Kumaran}}, \bibinfo {author} {\bibfnamefont {Christopher}\ \bibnamefont
  {Summerfield}}, \ and\ \bibinfo {author} {\bibfnamefont {Matthew}\
  \bibnamefont {Botvinick}},\ }\bibfield  {title} {\enquote {\bibinfo {title}
  {Neuroscience-inspired artificial intelligence},}\ }\href {\doibase
  10.1016/j.neuron.2017.06.011} {\bibfield  {journal} {\bibinfo  {journal}
  {Neuron}\ }\textbf {\bibinfo {volume} {95}},\ \bibinfo {pages} {245--258}
  (\bibinfo {year} {2017})}\BibitemShut {NoStop}%
\bibitem [{\citenamefont {Fudenberg}\ and\ \citenamefont
  {Levine}(1998)}]{FudenbergLevine1998}%
  \BibitemOpen
  \bibfield  {author} {\bibinfo {author} {\bibfnamefont {Drew}\ \bibnamefont
  {Fudenberg}}\ and\ \bibinfo {author} {\bibfnamefont {David~K}\ \bibnamefont
  {Levine}},\ }\href@noop {} {\emph {\bibinfo {title} {The Theory of Learning
  in Games}}}\ (\bibinfo  {publisher} {MIT Press},\ \bibinfo {year}
  {1998})\BibitemShut {NoStop}%
\bibitem [{\citenamefont {Roth}\ and\ \citenamefont
  {Erev}(1995)}]{RothErev1995}%
  \BibitemOpen
  \bibfield  {author} {\bibinfo {author} {\bibfnamefont {Alvin~E}\ \bibnamefont
  {Roth}}\ and\ \bibinfo {author} {\bibfnamefont {Ido}\ \bibnamefont {Erev}},\
  }\bibfield  {title} {\enquote {\bibinfo {title} {Learning in extensive-form
  games: Experimental data and simple dynamic models in the intermediate
  term},}\ }\href {\doibase 10.1016/s0899-8256(05)80020-x} {\bibfield
  {journal} {\bibinfo  {journal} {Games and Economic Behavior}\ }\textbf
  {\bibinfo {volume} {8}},\ \bibinfo {pages} {164--212} (\bibinfo {year}
  {1995})}\BibitemShut {NoStop}%
\bibitem [{\citenamefont {Erev}\ and\ \citenamefont
  {Roth}(1998)}]{ErevRoth1998}%
  \BibitemOpen
  \bibfield  {author} {\bibinfo {author} {\bibfnamefont {Ido}\ \bibnamefont
  {Erev}}\ and\ \bibinfo {author} {\bibfnamefont {Alvin~E}\ \bibnamefont
  {Roth}},\ }\bibfield  {title} {\enquote {\bibinfo {title} {Predicting how
  people play games: Reinforcement learning in experimental games with unique,
  mixed strategy equilibria},}\ }\href@noop {} {\bibfield  {journal} {\bibinfo
  {journal} {The American Economic Review}\ }\textbf {\bibinfo {volume} {88}},\
  \bibinfo {pages} {848--881} (\bibinfo {year} {1998})}\BibitemShut {NoStop}%
\bibitem [{\citenamefont {Camerer}\ and\ \citenamefont
  {Ho}(1999)}]{CamererHo1999}%
  \BibitemOpen
  \bibfield  {author} {\bibinfo {author} {\bibfnamefont {Colin}\ \bibnamefont
  {Camerer}}\ and\ \bibinfo {author} {\bibfnamefont {Teck~Hua}\ \bibnamefont
  {Ho}},\ }\bibfield  {title} {\enquote {\bibinfo {title} {Experience-weighted
  attraction learning in normal form games},}\ }\href {\doibase
  10.1111/1468-0262.00054} {\bibfield  {journal} {\bibinfo  {journal}
  {Econometrica}\ }\textbf {\bibinfo {volume} {67}},\ \bibinfo {pages}
  {827--874} (\bibinfo {year} {1999})}\BibitemShut {NoStop}%
\bibitem [{\citenamefont {Camerer}(2003)}]{Camerer2003}%
  \BibitemOpen
  \bibfield  {author} {\bibinfo {author} {\bibfnamefont {Colin~F}\ \bibnamefont
  {Camerer}},\ }\href@noop {} {\emph {\bibinfo {title} {Behavioral Game Theory:
  Experiments in Strategic Interaction}}}\ (\bibinfo  {publisher} {Princeton
  University Press},\ \bibinfo {year} {2003})\BibitemShut {NoStop}%
\bibitem [{\citenamefont {Arthur}(1993)}]{Arthur1991}%
  \BibitemOpen
  \bibfield  {author} {\bibinfo {author} {\bibfnamefont {W~Brian}\ \bibnamefont
  {Arthur}},\ }\bibfield  {title} {\enquote {\bibinfo {title} {On designing
  economic agents that behave like human agents},}\ }\href {\doibase
  10.1007/bf01199986} {\bibfield  {journal} {\bibinfo  {journal} {Journal of
  Evolutionary Economics}\ }\textbf {\bibinfo {volume} {3}},\ \bibinfo {pages}
  {1--22} (\bibinfo {year} {1993})}\BibitemShut {NoStop}%
\bibitem [{\citenamefont {Arthur}(1999)}]{Arthur1999}%
  \BibitemOpen
  \bibfield  {author} {\bibinfo {author} {\bibfnamefont {W~Brian}\ \bibnamefont
  {Arthur}},\ }\bibfield  {title} {\enquote {\bibinfo {title} {Complexity and
  the economy},}\ }\href {\doibase 10.1126/science.284.5411.107} {\bibfield
  {journal} {\bibinfo  {journal} {Science}\ }\textbf {\bibinfo {volume}
  {284}},\ \bibinfo {pages} {107--109} (\bibinfo {year} {1999})}\BibitemShut
  {NoStop}%
\bibitem [{\citenamefont {Macy}\ and\ \citenamefont
  {Flache}(2002)}]{MacyFlache2002}%
  \BibitemOpen
  \bibfield  {author} {\bibinfo {author} {\bibfnamefont {Michael~W}\
  \bibnamefont {Macy}}\ and\ \bibinfo {author} {\bibfnamefont {Andreas}\
  \bibnamefont {Flache}},\ }\bibfield  {title} {\enquote {\bibinfo {title}
  {Learning dynamics in social dilemmas},}\ }\href {\doibase
  10.1073/pnas.092080099} {\bibfield  {journal} {\bibinfo  {journal}
  {Proceedings of the National Academy of Sciences}\ }\textbf {\bibinfo
  {volume} {99}},\ \bibinfo {pages} {7229--7236} (\bibinfo {year}
  {2002})}\BibitemShut {NoStop}%
\bibitem [{\citenamefont {Cross}(1973)}]{Cross1973}%
  \BibitemOpen
  \bibfield  {author} {\bibinfo {author} {\bibfnamefont {John~G.}\ \bibnamefont
  {Cross}},\ }\bibfield  {title} {\enquote {\bibinfo {title} {A stochastic
  learning model of economic behavior},}\ }\href {\doibase 10.2307/1882186}
  {\bibfield  {journal} {\bibinfo  {journal} {The Quarterly Journal of
  Economics}\ }\textbf {\bibinfo {volume} {87}},\ \bibinfo {pages} {239}
  (\bibinfo {year} {1973})}\BibitemShut {NoStop}%
\bibitem [{\citenamefont {B{\"o}rgers}\ and\ \citenamefont
  {Sarin}(1997)}]{BoergersSarin1997}%
  \BibitemOpen
  \bibfield  {author} {\bibinfo {author} {\bibfnamefont {Tilman}\ \bibnamefont
  {B{\"o}rgers}}\ and\ \bibinfo {author} {\bibfnamefont {Rajiv}\ \bibnamefont
  {Sarin}},\ }\bibfield  {title} {\enquote {\bibinfo {title} {Learning through
  reinforcement and replicator dynamics},}\ }\href {\doibase
  10.1006/jeth.1997.2319} {\bibfield  {journal} {\bibinfo  {journal} {Journal
  of Economic Theory}\ }\textbf {\bibinfo {volume} {77}},\ \bibinfo {pages}
  {1--14} (\bibinfo {year} {1997})}\BibitemShut {NoStop}%
\bibitem [{\citenamefont {Marsili}\ \emph {et~al.}(2000)\citenamefont
  {Marsili}, \citenamefont {Challet},\ and\ \citenamefont
  {Zecchina}}]{MarsiliEtAl2000}%
  \BibitemOpen
  \bibfield  {author} {\bibinfo {author} {\bibfnamefont {Matteo}\ \bibnamefont
  {Marsili}}, \bibinfo {author} {\bibfnamefont {Damien}\ \bibnamefont
  {Challet}}, \ and\ \bibinfo {author} {\bibfnamefont {Riccardo}\ \bibnamefont
  {Zecchina}},\ }\bibfield  {title} {\enquote {\bibinfo {title} {Exact solution
  of a modified el farol's bar problem: Efficiency and the role of market
  impact},}\ }\href {\doibase 10.1016/s0378-4371(99)00610-x} {\bibfield
  {journal} {\bibinfo  {journal} {Physica A: Statistical Mechanics and its
  Applications}\ }\textbf {\bibinfo {volume} {280}},\ \bibinfo {pages}
  {522--553} (\bibinfo {year} {2000})}\BibitemShut {NoStop}%
\bibitem [{\citenamefont {Sato}\ \emph {et~al.}(2002)\citenamefont {Sato},
  \citenamefont {Akiyama},\ and\ \citenamefont {Farmer}}]{SatoEtAl2002}%
  \BibitemOpen
  \bibfield  {author} {\bibinfo {author} {\bibfnamefont {Yuzuru}\ \bibnamefont
  {Sato}}, \bibinfo {author} {\bibfnamefont {Eizo}\ \bibnamefont {Akiyama}}, \
  and\ \bibinfo {author} {\bibfnamefont {J~Doyne}\ \bibnamefont {Farmer}},\
  }\bibfield  {title} {\enquote {\bibinfo {title} {Chaos in learning a simple
  two-person game},}\ }\href {\doibase 10.1073/pnas.032086299} {\bibfield
  {journal} {\bibinfo  {journal} {Proceedings of the National Academy of
  Sciences}\ }\textbf {\bibinfo {volume} {99}},\ \bibinfo {pages} {4748--4751}
  (\bibinfo {year} {2002})}\BibitemShut {NoStop}%
\bibitem [{\citenamefont {Sato}\ and\ \citenamefont
  {Crutchfield}(2003)}]{SatoCrutchfield2003}%
  \BibitemOpen
  \bibfield  {author} {\bibinfo {author} {\bibfnamefont {Yuzuru}\ \bibnamefont
  {Sato}}\ and\ \bibinfo {author} {\bibfnamefont {James~P}\ \bibnamefont
  {Crutchfield}},\ }\bibfield  {title} {\enquote {\bibinfo {title} {Coupled
  replicator equations for the dynamics of learning in multiagent systems},}\
  }\href {\doibase 10.1103/physreve.67.015206} {\bibfield  {journal} {\bibinfo
  {journal} {Physical Review E}\ }\textbf {\bibinfo {volume} {67}} (\bibinfo
  {year} {2003}),\ 10.1103/physreve.67.015206}\BibitemShut {NoStop}%
\bibitem [{\citenamefont {Sato}\ \emph {et~al.}(2005)\citenamefont {Sato},
  \citenamefont {Akiyama},\ and\ \citenamefont {Crutchfield}}]{SatoEtAl2005}%
  \BibitemOpen
  \bibfield  {author} {\bibinfo {author} {\bibfnamefont {Yuzuru}\ \bibnamefont
  {Sato}}, \bibinfo {author} {\bibfnamefont {Eizo}\ \bibnamefont {Akiyama}}, \
  and\ \bibinfo {author} {\bibfnamefont {James~P}\ \bibnamefont
  {Crutchfield}},\ }\bibfield  {title} {\enquote {\bibinfo {title} {Stability
  and diversity in collective adaptation},}\ }\href {\doibase
  10.1016/j.physd.2005.06.031} {\bibfield  {journal} {\bibinfo  {journal}
  {Physica D: Nonlinear Phenomena}\ }\textbf {\bibinfo {volume} {210}},\
  \bibinfo {pages} {21--57} (\bibinfo {year} {2005})}\BibitemShut {NoStop}%
\bibitem [{\citenamefont {Galla}(2009)}]{Galla2009}%
  \BibitemOpen
  \bibfield  {author} {\bibinfo {author} {\bibfnamefont {Tobias}\ \bibnamefont
  {Galla}},\ }\bibfield  {title} {\enquote {\bibinfo {title} {Intrinsic noise
  in game dynamical learning},}\ }\href {\doibase
  10.1103/physrevlett.103.198702} {\bibfield  {journal} {\bibinfo  {journal}
  {Physical Review Letters}\ }\textbf {\bibinfo {volume} {103}} (\bibinfo
  {year} {2009}),\ 10.1103/physrevlett.103.198702}\BibitemShut {NoStop}%
\bibitem [{\citenamefont {Galla}(2011)}]{Galla2011}%
  \BibitemOpen
  \bibfield  {author} {\bibinfo {author} {\bibfnamefont {Tobias}\ \bibnamefont
  {Galla}},\ }\bibfield  {title} {\enquote {\bibinfo {title} {Cycles of
  cooperation and defection in imperfect learning},}\ }\href {\doibase
  10.1088/1742-5468/2011/08/p08007} {\bibfield  {journal} {\bibinfo  {journal}
  {Journal of Statistical Mechanics: Theory and Experiment}\ }\textbf {\bibinfo
  {volume} {2011}},\ \bibinfo {pages} {P08007} (\bibinfo {year}
  {2011})}\BibitemShut {NoStop}%
\bibitem [{\citenamefont {Bladon}\ and\ \citenamefont
  {Galla}(2011)}]{BladonGalla2011}%
  \BibitemOpen
  \bibfield  {author} {\bibinfo {author} {\bibfnamefont {Alex~J.}\ \bibnamefont
  {Bladon}}\ and\ \bibinfo {author} {\bibfnamefont {Tobias}\ \bibnamefont
  {Galla}},\ }\bibfield  {title} {\enquote {\bibinfo {title} {Learning dynamics
  in public goods games},}\ }\href {\doibase 10.1103/physreve.84.041132}
  {\bibfield  {journal} {\bibinfo  {journal} {Physical Review E}\ }\textbf
  {\bibinfo {volume} {84}} (\bibinfo {year} {2011}),\
  10.1103/physreve.84.041132}\BibitemShut {NoStop}%
\bibitem [{\citenamefont {Realpe-Gomez}\ \emph {et~al.}(2012)\citenamefont
  {Realpe-Gomez}, \citenamefont {Szczesny}, \citenamefont {Dall'Asta},\ and\
  \citenamefont {Galla}}]{RealpeGomezEtAl2012}%
  \BibitemOpen
  \bibfield  {author} {\bibinfo {author} {\bibfnamefont {John}\ \bibnamefont
  {Realpe-Gomez}}, \bibinfo {author} {\bibfnamefont {Bartosz}\ \bibnamefont
  {Szczesny}}, \bibinfo {author} {\bibfnamefont {Luca}\ \bibnamefont
  {Dall'Asta}}, \ and\ \bibinfo {author} {\bibfnamefont {Tobias}\ \bibnamefont
  {Galla}},\ }\bibfield  {title} {\enquote {\bibinfo {title} {Fixation and
  escape times in stochastic game learning},}\ }\href {\doibase
  10.1088/1742-5468/2012/10/p10022} {\bibfield  {journal} {\bibinfo  {journal}
  {Journal of Statistical Mechanics: Theory and Experiment}\ }\textbf {\bibinfo
  {volume} {2012}},\ \bibinfo {pages} {P10022} (\bibinfo {year}
  {2012})}\BibitemShut {NoStop}%
\bibitem [{\citenamefont {Sanders}\ \emph {et~al.}(2012)\citenamefont
  {Sanders}, \citenamefont {Galla},\ and\ \citenamefont
  {Shapiro}}]{SandersEtAl2012}%
  \BibitemOpen
  \bibfield  {author} {\bibinfo {author} {\bibfnamefont {James~BT}\
  \bibnamefont {Sanders}}, \bibinfo {author} {\bibfnamefont {Tobias}\
  \bibnamefont {Galla}}, \ and\ \bibinfo {author} {\bibfnamefont {Jonathan~L}\
  \bibnamefont {Shapiro}},\ }\bibfield  {title} {\enquote {\bibinfo {title}
  {Effects of noise on convergent game-learning dynamics},}\ }\href {\doibase
  10.1088/1751-8113/45/10/105001} {\bibfield  {journal} {\bibinfo  {journal}
  {Journal of Physics A: Mathematical and Theoretical}\ }\textbf {\bibinfo
  {volume} {45}},\ \bibinfo {pages} {105001} (\bibinfo {year}
  {2012})}\BibitemShut {NoStop}%
\bibitem [{\citenamefont {Galla}\ and\ \citenamefont
  {Farmer}(2013)}]{GallaFarmer2013}%
  \BibitemOpen
  \bibfield  {author} {\bibinfo {author} {\bibfnamefont {Tobias}\ \bibnamefont
  {Galla}}\ and\ \bibinfo {author} {\bibfnamefont {J.~Doyne}\ \bibnamefont
  {Farmer}},\ }\bibfield  {title} {\enquote {\bibinfo {title} {Complex dynamics
  in learning complicated games},}\ }\href {\doibase 10.1073/pnas.1109672110}
  {\bibfield  {journal} {\bibinfo  {journal} {Proceedings of the National
  Academy of Sciences}\ }\textbf {\bibinfo {volume} {110}},\ \bibinfo {pages}
  {1232--1236} (\bibinfo {year} {2013})}\BibitemShut {NoStop}%
\bibitem [{\citenamefont {Aloric}\ \emph {et~al.}(2016)\citenamefont {Aloric},
  \citenamefont {Sollich}, \citenamefont {McBurney},\ and\ \citenamefont
  {Galla}}]{AloricEtAl2016}%
  \BibitemOpen
  \bibfield  {author} {\bibinfo {author} {\bibfnamefont {Aleksandra}\
  \bibnamefont {Aloric}}, \bibinfo {author} {\bibfnamefont {Peter}\
  \bibnamefont {Sollich}}, \bibinfo {author} {\bibfnamefont {Peter}\
  \bibnamefont {McBurney}}, \ and\ \bibinfo {author} {\bibfnamefont {Tobias}\
  \bibnamefont {Galla}},\ }\bibfield  {title} {\enquote {\bibinfo {title}
  {{E}mergence of {C}ooperative {L}ong-{T}erm {M}arket {L}oyalty in {D}ouble
  {A}uction {M}arkets},}\ }\href {\doibase
  https://doi.org/10.1371/journal.pone.0154606} {\bibfield  {journal} {\bibinfo
   {journal} {PloS ONE}\ }\textbf {\bibinfo {volume} {11}},\ \bibinfo {pages}
  {e0154606} (\bibinfo {year} {2016})}\BibitemShut {NoStop}%
\bibitem [{\citenamefont {Tuyls}\ \emph {et~al.}(2003)\citenamefont {Tuyls},
  \citenamefont {Verbeeck},\ and\ \citenamefont {Lenaerts}}]{TuylsEtAl2003}%
  \BibitemOpen
  \bibfield  {author} {\bibinfo {author} {\bibfnamefont {Karl}\ \bibnamefont
  {Tuyls}}, \bibinfo {author} {\bibfnamefont {Katja}\ \bibnamefont {Verbeeck}},
  \ and\ \bibinfo {author} {\bibfnamefont {Tom}\ \bibnamefont {Lenaerts}},\
  }\bibfield  {title} {\enquote {\bibinfo {title} {A selection-mutation model
  for q-learning in multi-agent systems},}\ }in\ \href {\doibase
  10.1145/860575.860687} {\emph {\bibinfo {booktitle} {Proceedings of the
  Second International Joint Conference on Autonomous Agents and Multiagent
  Systems}}},\ \bibinfo {series and number} {AAMAS 2003}\ (\bibinfo {year}
  {2003})\ pp.\ \bibinfo {pages} {693--700}\BibitemShut {NoStop}%
\bibitem [{\citenamefont {Bloembergen}\ \emph {et~al.}(2015)\citenamefont
  {Bloembergen}, \citenamefont {Tuyls}, \citenamefont {Hennes},\ and\
  \citenamefont {Kaisers}}]{BloembergenEtAl2015}%
  \BibitemOpen
  \bibfield  {author} {\bibinfo {author} {\bibfnamefont {Daan}\ \bibnamefont
  {Bloembergen}}, \bibinfo {author} {\bibfnamefont {Karl}\ \bibnamefont
  {Tuyls}}, \bibinfo {author} {\bibfnamefont {Daniel}\ \bibnamefont {Hennes}},
  \ and\ \bibinfo {author} {\bibfnamefont {Michael}\ \bibnamefont {Kaisers}},\
  }\bibfield  {title} {\enquote {\bibinfo {title} {Evolutionary dynamics of
  multi-agent learning: A survey},}\ }\href {\doibase 10.1613/jair.4818}
  {\bibfield  {journal} {\bibinfo  {journal} {Journal of Artificial
  Intelligence Research}\ }\textbf {\bibinfo {volume} {53}},\ \bibinfo {pages}
  {659--697} (\bibinfo {year} {2015})}\BibitemShut {NoStop}%
\bibitem [{\citenamefont {Tuyls}\ and\ \citenamefont
  {Now{\'e}}(2005)}]{TuylsNowe2005}%
  \BibitemOpen
  \bibfield  {author} {\bibinfo {author} {\bibfnamefont {Karl}\ \bibnamefont
  {Tuyls}}\ and\ \bibinfo {author} {\bibfnamefont {Ann}\ \bibnamefont
  {Now{\'e}}},\ }\bibfield  {title} {\enquote {\bibinfo {title} {Evolutionary
  game theory and multi-agent reinforcement learning},}\ }\href {\doibase
  10.1017/s026988890500041x} {\bibfield  {journal} {\bibinfo  {journal} {The
  Knowledge Engineering Review}\ }\textbf {\bibinfo {volume} {20}},\ \bibinfo
  {pages} {63--90} (\bibinfo {year} {2005})}\BibitemShut {NoStop}%
\bibitem [{\citenamefont {Tuyls}\ \emph {et~al.}(2006)\citenamefont {Tuyls},
  \citenamefont {Hoen},\ and\ \citenamefont {Vanschoenwinkel}}]{TuylsEtAl2006}%
  \BibitemOpen
  \bibfield  {author} {\bibinfo {author} {\bibfnamefont {Karl}\ \bibnamefont
  {Tuyls}}, \bibinfo {author} {\bibfnamefont {Pieter~Jan'T}\ \bibnamefont
  {Hoen}}, \ and\ \bibinfo {author} {\bibfnamefont {Bram}\ \bibnamefont
  {Vanschoenwinkel}},\ }\bibfield  {title} {\enquote {\bibinfo {title} {An
  evolutionary dynamical analysis of multi-agent learning in iterated games},}\
  }\href {\doibase 10.1007/s10458-005-3783-9} {\bibfield  {journal} {\bibinfo
  {journal} {Autonomous Agents and Multi-Agent Systems}\ }\textbf {\bibinfo
  {volume} {12}},\ \bibinfo {pages} {115--153} (\bibinfo {year}
  {2006})}\BibitemShut {NoStop}%
\bibitem [{\citenamefont {Tuyls}\ and\ \citenamefont
  {Parsons}(2007)}]{TuylsParsons2007}%
  \BibitemOpen
  \bibfield  {author} {\bibinfo {author} {\bibfnamefont {Karl}\ \bibnamefont
  {Tuyls}}\ and\ \bibinfo {author} {\bibfnamefont {Simon}\ \bibnamefont
  {Parsons}},\ }\bibfield  {title} {\enquote {\bibinfo {title} {What
  evolutionary game theory tells us about multiagent learning},}\ }\href
  {\doibase 10.1016/j.artint.2007.01.004} {\bibfield  {journal} {\bibinfo
  {journal} {Artificial Intelligence}\ }\textbf {\bibinfo {volume} {171}},\
  \bibinfo {pages} {406--416} (\bibinfo {year} {2007})}\BibitemShut {NoStop}%
\bibitem [{\citenamefont {Kaisers}\ and\ \citenamefont
  {Tuyls}(2010)}]{KaisersTuyls2010}%
  \BibitemOpen
  \bibfield  {author} {\bibinfo {author} {\bibfnamefont {Michael}\ \bibnamefont
  {Kaisers}}\ and\ \bibinfo {author} {\bibfnamefont {Karl}\ \bibnamefont
  {Tuyls}},\ }\bibfield  {title} {\enquote {\bibinfo {title} {Frequency
  adjusted multi-agent q-learning},}\ }in\ \href@noop {} {\emph {\bibinfo
  {booktitle} {Proceedings of the 9th International Conference on Autonomous
  Agents and Multiagent Systems: Volume 1}}},\ \bibinfo {series and number}
  {AAMAS 2010}\ (\bibinfo {year} {2010})\ pp.\ \bibinfo {pages}
  {309--315}\BibitemShut {NoStop}%
\bibitem [{\citenamefont {Hennes}\ \emph {et~al.}(2009)\citenamefont {Hennes},
  \citenamefont {Tuyls},\ and\ \citenamefont {Rauterberg}}]{HennesEtAl2009}%
  \BibitemOpen
  \bibfield  {author} {\bibinfo {author} {\bibfnamefont {Daniel}\ \bibnamefont
  {Hennes}}, \bibinfo {author} {\bibfnamefont {Karl}\ \bibnamefont {Tuyls}}, \
  and\ \bibinfo {author} {\bibfnamefont {Matthias}\ \bibnamefont
  {Rauterberg}},\ }\bibfield  {title} {\enquote {\bibinfo {title}
  {State-coupled replicator dynamics},}\ }in\ \href@noop {} {\emph {\bibinfo
  {booktitle} {Proceedings of the 8th International Conference on Autonomous
  Agents and Multiagent Systems}}},\ \bibinfo {series and number} {AAMAS 2009}\
  (\bibinfo {year} {2009})\ pp.\ \bibinfo {pages} {789--796}\BibitemShut
  {NoStop}%
\bibitem [{\citenamefont {Vrancx}\ \emph {et~al.}(2008)\citenamefont {Vrancx},
  \citenamefont {Tuyls},\ and\ \citenamefont {Westra}}]{VrancxEtAl2008}%
  \BibitemOpen
  \bibfield  {author} {\bibinfo {author} {\bibfnamefont {Peter}\ \bibnamefont
  {Vrancx}}, \bibinfo {author} {\bibfnamefont {Karl}\ \bibnamefont {Tuyls}}, \
  and\ \bibinfo {author} {\bibfnamefont {Ronald}\ \bibnamefont {Westra}},\
  }\bibfield  {title} {\enquote {\bibinfo {title} {Switching dynamics of
  multi-agent learning},}\ }in\ \href@noop {} {\emph {\bibinfo {booktitle}
  {Proceedings of the 7th International Joint Conference on Autonomous Agents
  and Multiagent systems}}},\ \bibinfo {series and number} {AAMAS 2008}\
  (\bibinfo {year} {2008})\ pp.\ \bibinfo {pages} {307--313}\BibitemShut
  {NoStop}%
\bibitem [{\citenamefont {Hennes}\ \emph {et~al.}(2010)\citenamefont {Hennes},
  \citenamefont {Kaisers},\ and\ \citenamefont {Tuyls}}]{HennesEtAl2010}%
  \BibitemOpen
  \bibfield  {author} {\bibinfo {author} {\bibfnamefont {Daniel}\ \bibnamefont
  {Hennes}}, \bibinfo {author} {\bibfnamefont {Michael}\ \bibnamefont
  {Kaisers}}, \ and\ \bibinfo {author} {\bibfnamefont {Karl}\ \bibnamefont
  {Tuyls}},\ }\bibfield  {title} {\enquote {\bibinfo {title} {Resq-learning in
  stochastic games},}\ }in\ \href@noop {} {\emph {\bibinfo {booktitle}
  {Proceedings of the Adaptive and Learning Agents Workshop}}},\ \bibinfo
  {series and number} {ALA 2010}\ (\bibinfo {year} {2010})\ pp.\ \bibinfo
  {pages} {8--15}\BibitemShut {NoStop}%
\bibitem [{\citenamefont {Shapley}(1953)}]{Shapley1953}%
  \BibitemOpen
  \bibfield  {author} {\bibinfo {author} {\bibfnamefont {Lloyd~S}\ \bibnamefont
  {Shapley}},\ }\bibfield  {title} {\enquote {\bibinfo {title} {Stochastic
  games},}\ }\href {\doibase 10.1073/pnas.39.10.1095} {\bibfield  {journal}
  {\bibinfo  {journal} {Proceedings of the National Academy of Sciences}\
  }\textbf {\bibinfo {volume} {39}},\ \bibinfo {pages} {1095--1100} (\bibinfo
  {year} {1953})}\BibitemShut {NoStop}%
\bibitem [{\citenamefont {Mertens}\ and\ \citenamefont
  {Neyman}(1981)}]{MertensNeyman1981}%
  \BibitemOpen
  \bibfield  {author} {\bibinfo {author} {\bibfnamefont {J-F}\ \bibnamefont
  {Mertens}}\ and\ \bibinfo {author} {\bibfnamefont {Abraham}\ \bibnamefont
  {Neyman}},\ }\bibfield  {title} {\enquote {\bibinfo {title} {Stochastic
  games},}\ }\href@noop {} {\bibfield  {journal} {\bibinfo  {journal}
  {International Journal of Game Theory}\ }\textbf {\bibinfo {volume} {10}},\
  \bibinfo {pages} {53--66} (\bibinfo {year} {1981})}\BibitemShut {NoStop}%
\bibitem [{\citenamefont {Akiyama}\ and\ \citenamefont
  {Kaneko}(2000)}]{AkiyamaKaneko2000}%
  \BibitemOpen
  \bibfield  {author} {\bibinfo {author} {\bibfnamefont {Eizo}\ \bibnamefont
  {Akiyama}}\ and\ \bibinfo {author} {\bibfnamefont {Kunihiko}\ \bibnamefont
  {Kaneko}},\ }\bibfield  {title} {\enquote {\bibinfo {title} {{Dynamical
  systems game theory and dynamics of games}},}\ }\href {\doibase
  10.1016/S0167-2789(00)00157-3} {\bibfield  {journal} {\bibinfo  {journal}
  {Physica D: Nonlinear Phenomena}\ }\textbf {\bibinfo {volume} {147}},\
  \bibinfo {pages} {221--258} (\bibinfo {year} {2000})}\BibitemShut {NoStop}%
\bibitem [{\citenamefont {Akiyama}\ and\ \citenamefont
  {Kaneko}(2002)}]{AkiyamaKaneko2002}%
  \BibitemOpen
  \bibfield  {author} {\bibinfo {author} {\bibfnamefont {Eizo}\ \bibnamefont
  {Akiyama}}\ and\ \bibinfo {author} {\bibfnamefont {Kunihiko}\ \bibnamefont
  {Kaneko}},\ }\bibfield  {title} {\enquote {\bibinfo {title} {{Dynamical
  systems game theory II: A new approach to the problem of the social
  dilemma}},}\ }\href {\doibase 10.1016/S0167-2789(02)00402-5} {\bibfield
  {journal} {\bibinfo  {journal} {Physica D: Nonlinear Phenomena}\ }\textbf
  {\bibinfo {volume} {167}},\ \bibinfo {pages} {36--71} (\bibinfo {year}
  {2002})}\BibitemShut {NoStop}%
\bibitem [{\citenamefont {Spaan}(2012)}]{Spaan2012}%
  \BibitemOpen
  \bibfield  {author} {\bibinfo {author} {\bibfnamefont {Matthijs T.~J.}\
  \bibnamefont {Spaan}},\ }\bibfield  {title} {\enquote {\bibinfo {title}
  {Partially observable markov decision processes},}\ }in\ \href {\doibase
  10.1007/978-3-642-27645-3_12} {\emph {\bibinfo {booktitle} {Reinforcement
  Learning}}}\ (\bibinfo  {publisher} {Springer Berlin Heidelberg},\ \bibinfo
  {year} {2012})\ pp.\ \bibinfo {pages} {387--414}\BibitemShut {NoStop}%
\bibitem [{\citenamefont {Oliehoek}(2012)}]{Oliehoek2012}%
  \BibitemOpen
  \bibfield  {author} {\bibinfo {author} {\bibfnamefont {Frans~A.}\
  \bibnamefont {Oliehoek}},\ }\bibfield  {title} {\enquote {\bibinfo {title}
  {Decentralized {POMDPs}},}\ }in\ \href {\doibase
  10.1007/978-3-642-27645-3_15} {\emph {\bibinfo {booktitle} {Adaptation,
  Learning, and Optimization}}}\ (\bibinfo  {publisher} {Springer Berlin
  Heidelberg},\ \bibinfo {year} {2012})\ pp.\ \bibinfo {pages}
  {471--503}\BibitemShut {NoStop}%
\bibitem [{\citenamefont {Bellman}(1957)}]{Bellman1957}%
  \BibitemOpen
  \bibfield  {author} {\bibinfo {author} {\bibfnamefont {Richard}\ \bibnamefont
  {Bellman}},\ }\bibfield  {title} {\enquote {\bibinfo {title} {A markovian
  decision process},}\ }\href {\doibase 10.1512/iumj.1957.6.56038} {\bibfield
  {journal} {\bibinfo  {journal} {Indiana University Mathematics Journal}\
  }\textbf {\bibinfo {volume} {6}},\ \bibinfo {pages} {679--684} (\bibinfo
  {year} {1957})}\BibitemShut {NoStop}%
\bibitem [{\citenamefont {Lange}\ \emph {et~al.}(2012)\citenamefont {Lange},
  \citenamefont {Gabel},\ and\ \citenamefont {Riedmiller}}]{LangeEtAl2012}%
  \BibitemOpen
  \bibfield  {author} {\bibinfo {author} {\bibfnamefont {Sascha}\ \bibnamefont
  {Lange}}, \bibinfo {author} {\bibfnamefont {Thomas}\ \bibnamefont {Gabel}}, \
  and\ \bibinfo {author} {\bibfnamefont {Martin}\ \bibnamefont {Riedmiller}},\
  }\bibfield  {title} {\enquote {\bibinfo {title} {Batch reinforcement
  learning},}\ }in\ \href {\doibase 10.1007/978-3-642-27645-3_2} {\emph
  {\bibinfo {booktitle} {Reinforcement Learning}}}\ (\bibinfo  {publisher}
  {Springer Berlin Heidelberg},\ \bibinfo {year} {2012})\ pp.\ \bibinfo {pages}
  {45--73}\BibitemShut {NoStop}%
\bibitem [{\citenamefont {Mnih}\ \emph {et~al.}(2015)\citenamefont {Mnih},
  \citenamefont {Kavukcuoglu}, \citenamefont {Silver}, \citenamefont {Rusu},
  \citenamefont {Veness}, \citenamefont {Bellemare}, \citenamefont {Graves},
  \citenamefont {Riedmiller}, \citenamefont {Fidjeland}, \citenamefont
  {Ostrovski}, \citenamefont {Petersen}, \citenamefont {Beattie}, \citenamefont
  {Sadik}, \citenamefont {Antonoglou}, \citenamefont {King}, \citenamefont
  {Kumaran}, \citenamefont {Wierstra}, \citenamefont {Legg},\ and\
  \citenamefont {Hassabis}}]{MnihEtAl2015}%
  \BibitemOpen
  \bibfield  {author} {\bibinfo {author} {\bibfnamefont {Volodymyr}\
  \bibnamefont {Mnih}}, \bibinfo {author} {\bibfnamefont {Koray}\ \bibnamefont
  {Kavukcuoglu}}, \bibinfo {author} {\bibfnamefont {David}\ \bibnamefont
  {Silver}}, \bibinfo {author} {\bibfnamefont {Andrei~A.}\ \bibnamefont
  {Rusu}}, \bibinfo {author} {\bibfnamefont {Joel}\ \bibnamefont {Veness}},
  \bibinfo {author} {\bibfnamefont {Marc~G.}\ \bibnamefont {Bellemare}},
  \bibinfo {author} {\bibfnamefont {Alex}\ \bibnamefont {Graves}}, \bibinfo
  {author} {\bibfnamefont {Martin}\ \bibnamefont {Riedmiller}}, \bibinfo
  {author} {\bibfnamefont {Andreas~K.}\ \bibnamefont {Fidjeland}}, \bibinfo
  {author} {\bibfnamefont {Georg}\ \bibnamefont {Ostrovski}}, \bibinfo {author}
  {\bibfnamefont {Stig}\ \bibnamefont {Petersen}}, \bibinfo {author}
  {\bibfnamefont {Charles}\ \bibnamefont {Beattie}}, \bibinfo {author}
  {\bibfnamefont {Amir}\ \bibnamefont {Sadik}}, \bibinfo {author}
  {\bibfnamefont {Ioannis}\ \bibnamefont {Antonoglou}}, \bibinfo {author}
  {\bibfnamefont {Helen}\ \bibnamefont {King}}, \bibinfo {author}
  {\bibfnamefont {Dharshan}\ \bibnamefont {Kumaran}}, \bibinfo {author}
  {\bibfnamefont {Daan}\ \bibnamefont {Wierstra}}, \bibinfo {author}
  {\bibfnamefont {Shane}\ \bibnamefont {Legg}}, \ and\ \bibinfo {author}
  {\bibfnamefont {Demis}\ \bibnamefont {Hassabis}},\ }\bibfield  {title}
  {\enquote {\bibinfo {title} {Human-level control through deep reinforcement
  learning},}\ }\href {\doibase 10.1038/nature14236} {\bibfield  {journal}
  {\bibinfo  {journal} {Nature}\ }\textbf {\bibinfo {volume} {518}},\ \bibinfo
  {pages} {529--533} (\bibinfo {year} {2015})}\BibitemShut {NoStop}%
\bibitem [{\citenamefont {Hilbe}\ \emph {et~al.}(2018)\citenamefont {Hilbe},
  \citenamefont {{\v{S}}imsa}, \citenamefont {Chatterjee},\ and\ \citenamefont
  {Nowak}}]{HilbeEtAl2018}%
  \BibitemOpen
  \bibfield  {author} {\bibinfo {author} {\bibfnamefont {Christian}\
  \bibnamefont {Hilbe}}, \bibinfo {author} {\bibfnamefont
  {{\v{S}}t{\v{e}}p{\'{a}}n}\ \bibnamefont {{\v{S}}imsa}}, \bibinfo {author}
  {\bibfnamefont {Krishnendu}\ \bibnamefont {Chatterjee}}, \ and\ \bibinfo
  {author} {\bibfnamefont {Martin~A.}\ \bibnamefont {Nowak}},\ }\bibfield
  {title} {\enquote {\bibinfo {title} {Evolution of cooperation in stochastic
  games},}\ }\href {\doibase 10.1038/s41586-018-0277-x} {\bibfield  {journal}
  {\bibinfo  {journal} {Nature}\ }\textbf {\bibinfo {volume} {559}},\ \bibinfo
  {pages} {246--249} (\bibinfo {year} {2018})}\BibitemShut {NoStop}%
\bibitem [{\citenamefont {Sandri}(1996)}]{Sandri1996}%
  \BibitemOpen
  \bibfield  {author} {\bibinfo {author} {\bibfnamefont {Marco}\ \bibnamefont
  {Sandri}},\ }\bibfield  {title} {\enquote {\bibinfo {title} {Numerical
  calculation of lyapunov exponents},}\ }\href@noop {} {\bibfield  {journal}
  {\bibinfo  {journal} {The Mathematica Journal}\ }\textbf {\bibinfo {volume}
  {6}},\ \bibinfo {pages} {78--84} (\bibinfo {year} {1996})}\BibitemShut
  {NoStop}%
\bibitem [{\citenamefont {Bandura}(1977)}]{Bandura1977}%
  \BibitemOpen
  \bibfield  {author} {\bibinfo {author} {\bibfnamefont {A.}~\bibnamefont
  {Bandura}},\ }\href@noop {} {\emph {\bibinfo {title} {Social learning
  Theory}}}\ (\bibinfo  {publisher} {Englewood Cliffs, NJ: Prentice Hall},\
  \bibinfo {year} {1977})\BibitemShut {NoStop}%
\bibitem [{\citenamefont {Smolla}\ \emph {et~al.}(2015)\citenamefont {Smolla},
  \citenamefont {Gilman}, \citenamefont {Galla},\ and\ \citenamefont
  {Shultz}}]{SmollaEtAl2015}%
  \BibitemOpen
  \bibfield  {author} {\bibinfo {author} {\bibfnamefont {Marco}\ \bibnamefont
  {Smolla}}, \bibinfo {author} {\bibfnamefont {R.~Tucker}\ \bibnamefont
  {Gilman}}, \bibinfo {author} {\bibfnamefont {Tobias}\ \bibnamefont {Galla}},
  \ and\ \bibinfo {author} {\bibfnamefont {Susanne}\ \bibnamefont {Shultz}},\
  }\bibfield  {title} {\enquote {\bibinfo {title} {Competition for resources
  can explain patterns of social and individual learning in nature},}\ }\href
  {\doibase 10.1098/rspb.2015.1405} {\bibfield  {journal} {\bibinfo  {journal}
  {Proceedings of the Royal Society B: Biological Sciences}\ }\textbf {\bibinfo
  {volume} {282}},\ \bibinfo {pages} {20151405} (\bibinfo {year}
  {2015})}\BibitemShut {NoStop}%
\bibitem [{\citenamefont {Barfuss}\ \emph {et~al.}(2017, P1)\citenamefont
  {Barfuss}, \citenamefont {Donges}, \citenamefont {Wiedermann},\ and\
  \citenamefont {Lucht}}]{BarfussEtAl2017}%
  \BibitemOpen
  \bibfield  {author} {\bibinfo {author} {\bibfnamefont {Wolfram}\ \bibnamefont
  {Barfuss}}, \bibinfo {author} {\bibfnamefont {Jonathan~F}\ \bibnamefont
  {Donges}}, \bibinfo {author} {\bibfnamefont {Marc}\ \bibnamefont
  {Wiedermann}}, \ and\ \bibinfo {author} {\bibfnamefont {Wolfgang}\
  \bibnamefont {Lucht}},\ }\bibfield  {title} {\enquote {\bibinfo {title}
  {Sustainable use of renewable resources in a stylized social--ecological
  network model under heterogeneous resource distribution},}\ }\href {\doibase
  10.5194/esd-8-255-2017} {\bibfield  {journal} {\bibinfo  {journal} {Earth
  System Dynamics}\ }\textbf {\bibinfo {volume} {8}},\ \bibinfo {pages}
  {255--264} (\bibinfo {year} {2017, P1})}\BibitemShut {NoStop}%
\bibitem [{\citenamefont {Banisch}\ and\ \citenamefont
  {Olbrich}(2018)}]{BanischOlbrich2018}%
  \BibitemOpen
  \bibfield  {author} {\bibinfo {author} {\bibfnamefont {S.}~\bibnamefont
  {Banisch}}\ and\ \bibinfo {author} {\bibfnamefont {E.}~\bibnamefont
  {Olbrich}},\ }\bibfield  {title} {\enquote {\bibinfo {title} {Opinion
  polarization by learning from social feedback},}\ }\href {\doibase
  10.1080/0022250X.2018.1517761} {\bibfield  {journal} {\bibinfo  {journal}
  {The Journal of Mathematical Sociology}\ }\textbf {\bibinfo {volume} {0}},\
  \bibinfo {pages} {1--28} (\bibinfo {year} {2018})}\BibitemShut {NoStop}%
\bibitem [{\citenamefont {Levin}(2013)}]{Levin2013}%
  \BibitemOpen
  \bibfield  {author} {\bibinfo {author} {\bibfnamefont {Simon}\ \bibnamefont
  {Levin}},\ }\bibfield  {title} {\enquote {\bibinfo {title} {The mathematics
  of sustainability},}\ }\href {\doibase 10.1090/noti982} {\bibfield  {journal}
  {\bibinfo  {journal} {Notices of the American Mathematical Society}\ }\textbf
  {\bibinfo {volume} {60}},\ \bibinfo {pages} {1} (\bibinfo {year}
  {2013})}\BibitemShut {NoStop}%
\bibitem [{\citenamefont {Donges}\ \emph {et~al.}(2017)\citenamefont {Donges},
  \citenamefont {Winkelmann}, \citenamefont {Lucht}, \citenamefont {Cornell},
  \citenamefont {Dyke}, \citenamefont {Rockstr{\"o}m}, \citenamefont
  {Heitzig},\ and\ \citenamefont {Schellnhuber}}]{DongesEtAl2017}%
  \BibitemOpen
  \bibfield  {author} {\bibinfo {author} {\bibfnamefont {Jonathan~F}\
  \bibnamefont {Donges}}, \bibinfo {author} {\bibfnamefont {Ricarda}\
  \bibnamefont {Winkelmann}}, \bibinfo {author} {\bibfnamefont {Wolfgang}\
  \bibnamefont {Lucht}}, \bibinfo {author} {\bibfnamefont {Sarah~E}\
  \bibnamefont {Cornell}}, \bibinfo {author} {\bibfnamefont {James~G}\
  \bibnamefont {Dyke}}, \bibinfo {author} {\bibfnamefont {Johan}\ \bibnamefont
  {Rockstr{\"o}m}}, \bibinfo {author} {\bibfnamefont {Jobst}\ \bibnamefont
  {Heitzig}}, \ and\ \bibinfo {author} {\bibfnamefont {Hans~Joachim}\
  \bibnamefont {Schellnhuber}},\ }\bibfield  {title} {\enquote {\bibinfo
  {title} {Closing the loop: Reconnecting human dynamics to earth system
  science},}\ }\href {\doibase 10.1177/2053019617725537} {\bibfield  {journal}
  {\bibinfo  {journal} {The Anthropocene Review}\ }\textbf {\bibinfo {volume}
  {4}},\ \bibinfo {pages} {151--157} (\bibinfo {year} {2017})}\BibitemShut
  {NoStop}%
\bibitem [{\citenamefont {Dawes}(1980)}]{Dawes1980}%
  \BibitemOpen
  \bibfield  {author} {\bibinfo {author} {\bibfnamefont {R~M}\ \bibnamefont
  {Dawes}},\ }\bibfield  {title} {\enquote {\bibinfo {title} {Social
  dilemmas},}\ }\href {\doibase 10.1146/annurev.ps.31.020180.001125} {\bibfield
   {journal} {\bibinfo  {journal} {Annual Review of Psychology}\ }\textbf
  {\bibinfo {volume} {31}},\ \bibinfo {pages} {169--193} (\bibinfo {year}
  {1980})}\BibitemShut {NoStop}%
\bibitem [{\citenamefont {Heitzig}\ \emph {et~al.}(2016)\citenamefont
  {Heitzig}, \citenamefont {Kittel}, \citenamefont {Donges},\ and\
  \citenamefont {Molkenthin}}]{HeitzigEtAl2016}%
  \BibitemOpen
  \bibfield  {author} {\bibinfo {author} {\bibfnamefont {J.}~\bibnamefont
  {Heitzig}}, \bibinfo {author} {\bibfnamefont {T.}~\bibnamefont {Kittel}},
  \bibinfo {author} {\bibfnamefont {J.~F.}\ \bibnamefont {Donges}}, \ and\
  \bibinfo {author} {\bibfnamefont {N.}~\bibnamefont {Molkenthin}},\ }\bibfield
   {title} {\enquote {\bibinfo {title} {Topology of sustainable management of
  dynamical systems with desirable states: from defining planetary boundaries
  to safe operating spaces in the earth system},}\ }\href {\doibase
  10.5194/esd-7-21-2016} {\bibfield  {journal} {\bibinfo  {journal} {Earth
  System Dynamics}\ }\textbf {\bibinfo {volume} {7}},\ \bibinfo {pages}
  {21--50} (\bibinfo {year} {2016})}\BibitemShut {NoStop}%
\bibitem [{\citenamefont {Lindkvist}\ and\ \citenamefont
  {Norberg}(2014)}]{LindkvistNorberg2014}%
  \BibitemOpen
  \bibfield  {author} {\bibinfo {author} {\bibfnamefont {Emilie}\ \bibnamefont
  {Lindkvist}}\ and\ \bibinfo {author} {\bibfnamefont {Jon}\ \bibnamefont
  {Norberg}},\ }\bibfield  {title} {\enquote {\bibinfo {title} {Modeling
  experiential learning: The challenges posed by threshold dynamics for
  sustainable renewable resource management},}\ }\href {\doibase
  10.1016/j.ecolecon.2014.04.018} {\bibfield  {journal} {\bibinfo  {journal}
  {Ecological Economics}\ }\textbf {\bibinfo {volume} {104}},\ \bibinfo {pages}
  {107--118} (\bibinfo {year} {2014})}\BibitemShut {NoStop}%
\bibitem [{\citenamefont {Schill}\ \emph {et~al.}(2015)\citenamefont {Schill},
  \citenamefont {Lindahl},\ and\ \citenamefont {Cr{\'e}pin}}]{SchillEtAl2015}%
  \BibitemOpen
  \bibfield  {author} {\bibinfo {author} {\bibfnamefont {Caroline}\
  \bibnamefont {Schill}}, \bibinfo {author} {\bibfnamefont {Therese}\
  \bibnamefont {Lindahl}}, \ and\ \bibinfo {author} {\bibfnamefont
  {Anne-Sophie}\ \bibnamefont {Cr{\'e}pin}},\ }\bibfield  {title} {\enquote
  {\bibinfo {title} {Collective action and the risk of ecosystem regime shifts:
  insights from a laboratory experiment},}\ }\href {\doibase
  10.5751/es-07318-200148} {\bibfield  {journal} {\bibinfo  {journal} {Ecology
  and Society}\ }\textbf {\bibinfo {volume} {20}} (\bibinfo {year} {2015}),\
  10.5751/es-07318-200148}\BibitemShut {NoStop}%
\bibitem [{\citenamefont {Milinski}\ \emph {et~al.}(2008)\citenamefont
  {Milinski}, \citenamefont {Sommerfeld}, \citenamefont {Krambeck},
  \citenamefont {Reed},\ and\ \citenamefont {Marotzke}}]{MilinskiEtAl2008}%
  \BibitemOpen
  \bibfield  {author} {\bibinfo {author} {\bibfnamefont {Manfred}\ \bibnamefont
  {Milinski}}, \bibinfo {author} {\bibfnamefont {Ralf~D}\ \bibnamefont
  {Sommerfeld}}, \bibinfo {author} {\bibfnamefont {Hans-J{\"u}rgen}\
  \bibnamefont {Krambeck}}, \bibinfo {author} {\bibfnamefont {Floyd~A}\
  \bibnamefont {Reed}}, \ and\ \bibinfo {author} {\bibfnamefont {Jochem}\
  \bibnamefont {Marotzke}},\ }\bibfield  {title} {\enquote {\bibinfo {title}
  {The collective-risk social dilemma and the prevention of simulated dangerous
  climate change},}\ }\href {\doibase 10.1073/pnas.0709546105} {\bibfield
  {journal} {\bibinfo  {journal} {Proceedings of the National Academy of
  Sciences}\ }\textbf {\bibinfo {volume} {105}},\ \bibinfo {pages} {2291--2294}
  (\bibinfo {year} {2008})}\BibitemShut {NoStop}%
\bibitem [{\citenamefont {Barrett}\ and\ \citenamefont
  {Dannenberg}(2012)}]{BarrettDannenberg2012}%
  \BibitemOpen
  \bibfield  {author} {\bibinfo {author} {\bibfnamefont {S.}~\bibnamefont
  {Barrett}}\ and\ \bibinfo {author} {\bibfnamefont {A.}~\bibnamefont
  {Dannenberg}},\ }\bibfield  {title} {\enquote {\bibinfo {title} {Climate
  negotiations under scientific uncertainty},}\ }\href {\doibase
  10.1073/pnas.1208417109} {\bibfield  {journal} {\bibinfo  {journal}
  {Proceedings of the National Academy of Sciences}\ }\textbf {\bibinfo
  {volume} {109}},\ \bibinfo {pages} {17372--17376} (\bibinfo {year}
  {2012})}\BibitemShut {NoStop}%
\bibitem [{\citenamefont {Barfuss}\ \emph {et~al.}(2018, P4)\citenamefont
  {Barfuss}, \citenamefont {Donges}, \citenamefont {Lade},\ and\ \citenamefont
  {Kurths}}]{BarfussEtAl2018}%
  \BibitemOpen
  \bibfield  {author} {\bibinfo {author} {\bibfnamefont {Wolfram}\ \bibnamefont
  {Barfuss}}, \bibinfo {author} {\bibfnamefont {Jonathan~F}\ \bibnamefont
  {Donges}}, \bibinfo {author} {\bibfnamefont {Steven~J}\ \bibnamefont {Lade}},
  \ and\ \bibinfo {author} {\bibfnamefont {J{\"u}rgen}\ \bibnamefont
  {Kurths}},\ }\bibfield  {title} {\enquote {\bibinfo {title} {When
  optimization for governing human-environment tipping elements is neither
  sustainable nor safe},}\ }\href {\doibase 10.1038/s41467-018-04738-z}
  {\bibfield  {journal} {\bibinfo  {journal} {Nature communications}\ }\textbf
  {\bibinfo {volume} {9}},\ \bibinfo {pages} {2354} (\bibinfo {year} {2018,
  P4})}\BibitemShut {NoStop}%
\bibitem [{\citenamefont {Donges}\ and\ \citenamefont {Barfuss}(2017,
  P2)}]{DongesBarfuss2017}%
  \BibitemOpen
  \bibfield  {author} {\bibinfo {author} {\bibfnamefont {Jonathan~F.}\
  \bibnamefont {Donges}}\ and\ \bibinfo {author} {\bibfnamefont {Wolfram}\
  \bibnamefont {Barfuss}},\ }\bibfield  {title} {\enquote {\bibinfo {title}
  {From math to metaphors and back again: Social-ecological resilience from a
  multi-agent-environment perspective},}\ }\href {\doibase
  10.14512/gaia.26.s1.5} {\bibfield  {journal} {\bibinfo  {journal} {{GAIA} -
  Ecological Perspectives for Science and Society}\ }\textbf {\bibinfo {volume}
  {26}},\ \bibinfo {pages} {182--190} (\bibinfo {year} {2017, P2})}\BibitemShut
  {NoStop}%
\end{thebibliography}
